\documentclass[reprint, amsmath, amssymb, superscriptaddress, pra]{revtex4-2}
\usepackage{graphicx}
\usepackage{hyperref}
\usepackage{xcolor}
\usepackage{bbm}

\begin{document}
\title{Quantum Fourier Transform in Oscillating Modes}
\author{Qi-Ming Chen}
\affiliation{Walther-Mei{\ss}ner-Institut, Bayerische Akademie der Wissenschaften, 85748 Garching, Germany}
\affiliation{Physik-Department, Technische Universit{\"a}t M{\"u}nchen, 85748 Garching, Germany}

\author{Frank Deppe}
\email{frank.deppe@wmi.badw.de}
\affiliation{Walther-Mei{\ss}ner-Institut, Bayerische Akademie der Wissenschaften, 85748 Garching, Germany}
\affiliation{Physik-Department, Technische Universit{\"a}t M{\"u}nchen, 85748 Garching, Germany}
\affiliation{Munich Center for Quantum Science and Technology (MCQST), Schellingstr. 4, 80799 Munich, Germany}

\author{Re-Bing Wu}
\affiliation{Department of Automation, Tsinghua University, Beijing 100084, China}
\affiliation{Beijing National Research Center for Information Science and Technology, Beijing 100084, China}

\author{Luyan Sun}
\affiliation{Center for Quantum Information, Institute for Interdisciplinary Information Sciences, Tsinghua University, Beijing 100084, China}

\author{Yu-xi Liu}
\affiliation{Beijing National Research Center for Information Science and Technology, Beijing 100084, China}
\affiliation{Institute of Microelectronics, Tsinghua University, Beijing 100084, China}

\author{Yuki Nojiri}
\author{Stefan Pogorzalek}
\author{Michael Renger}
\affiliation{Walther-Mei{\ss}ner-Institut, Bayerische Akademie der Wissenschaften, 85748 Garching, Germany}
\affiliation{Physik-Department, Technische Universit{\"a}t M{\"u}nchen, 85748 Garching, Germany}

\author{Matti Partanen}
\affiliation{Walther-Mei{\ss}ner-Institut, Bayerische Akademie der Wissenschaften, 85748 Garching, Germany}

\author{Kirill G. Fedorov}
\affiliation{Walther-Mei{\ss}ner-Institut, Bayerische Akademie der Wissenschaften, 85748 Garching, Germany}
\affiliation{Physik-Department, Technische Universit{\"a}t M{\"u}nchen, 85748 Garching, Germany}

\author{Achim Marx}
\affiliation{Walther-Mei{\ss}ner-Institut, Bayerische Akademie der Wissenschaften, 85748 Garching, Germany}

\author{Rudolf Gross}
\email{rudolf.gross@wmi.badw.de}
\affiliation{Walther-Mei{\ss}ner-Institut, Bayerische Akademie der Wissenschaften, 85748 Garching, Germany}
\affiliation{Physik-Department, Technische Universit{\"a}t M{\"u}nchen, 85748 Garching, Germany}
\affiliation{Munich Center for Quantum Science and Technology (MCQST), Schellingstr. 4, 80799 Munich, Germany}

\date{\today}

\begin{abstract}
	Quantum Fourier transform (QFT) is a key ingredient of many quantum algorithms where a considerable amount of ancilla qubits and gates are often needed to form a Hilbert space large enough for high-precision results. Qubit recycling reduces the number of ancilla qubits to one but imposes the requirement of repeated measurements and feedforward within the coherence time of the qubits. Moreover, recycling only applies to certain cases where QFT can be carried out in a semi-classical way. Here, we report a novel approach based on two harmonic resonators which form a high-dimensional Hilbert space for the realization of QFT. By employing the \textit{all-resonant} and \textit{perfect} state-transfer methods, we develop a protocol that transfers an unknown multi-qubit state to one resonator. QFT is performed by the free evolution of the two resonators with a cross-Kerr interaction. Then, the fully-quantum result can be localized in the second resonator by a projective measurement. Qualitative analysis shows that a $2^{10}$-dimensional QFT can be realized in current superconducting quantum circuits which paves the way for implementing various quantum algorithms in the noisy intermediate-scale quantum (NISQ) era.
\end{abstract}

\maketitle

\section{Introduction}
In order to fully explore the quantum advantage in solving computational problems \cite{Neill2018, Arute2019}, various quantum algorithms have been proposed during the last decades \cite{Nielsen2010}. Among them, quantum Fourier transform (QFT) attracts a particularly high interest since it is the cornerstone of a considerable number of quantum algorithms. Prominent examples are the phase estimation algorithm, as well as its further applications in order-finding and factorization problems \cite{Nielsen2010}. To implement a $q$-dimensional QFT in a quantum circuit, one requires $n=\log_2(q)$ qubits as well as $\mathcal{O}\left(n^2\right)$ high-fidelity single or two-qubit gates \cite{Nielsen2010}. In typical applications of QFT, such as Shor's factorization algorithm \cite{Shor1994, *Shor1999}, these qubits are often required for implementing QFT tasks only, besides those carrying the actual quantum information in the algorithm. Due to these reasons, there have been only small-scale implementations of QFT, with Hilbert spaces spanned by up to three qubits, to the best of our knowledge \cite{Weinstein2001, Vandersypen2001, Politi2009, Lucero2012, Cai2013, Pan2014, Zheng2017}. Towards higher dimensions, a digital-analog approach, on the one hand, is proposed to improve the gate fidelity in practical implementations \cite{Martin2020}. On the other hand, qubit recycling has been used to reduce the number of ancilla qubits to one in certain cases \cite{Griffiths1996, Parker2000, Lu2007, Lanyon2007, Martin-Lopez2012, Zhou2013, Smolin2013, Monz2016}. However, the latter requires $n$ high-precision measurements combined with real-time feedforward within the coherence time of the system, which is technically difficult for large $q$. More importantly, it is a semi-classical method, which is not directly applicable to the algorithms where quantum coherence in the result should be kept for further quantum information processing, e.g. the Harrow-Hassidim-Lloyd (HHL) scheme for solving linear equations \cite{Harrow2009}. Other examples may include period-founding, discrete logarithms, and hidden subgroup problems \cite{Nielsen2010}.

\begin{figure}
  \centering
  \includegraphics[width=\columnwidth]{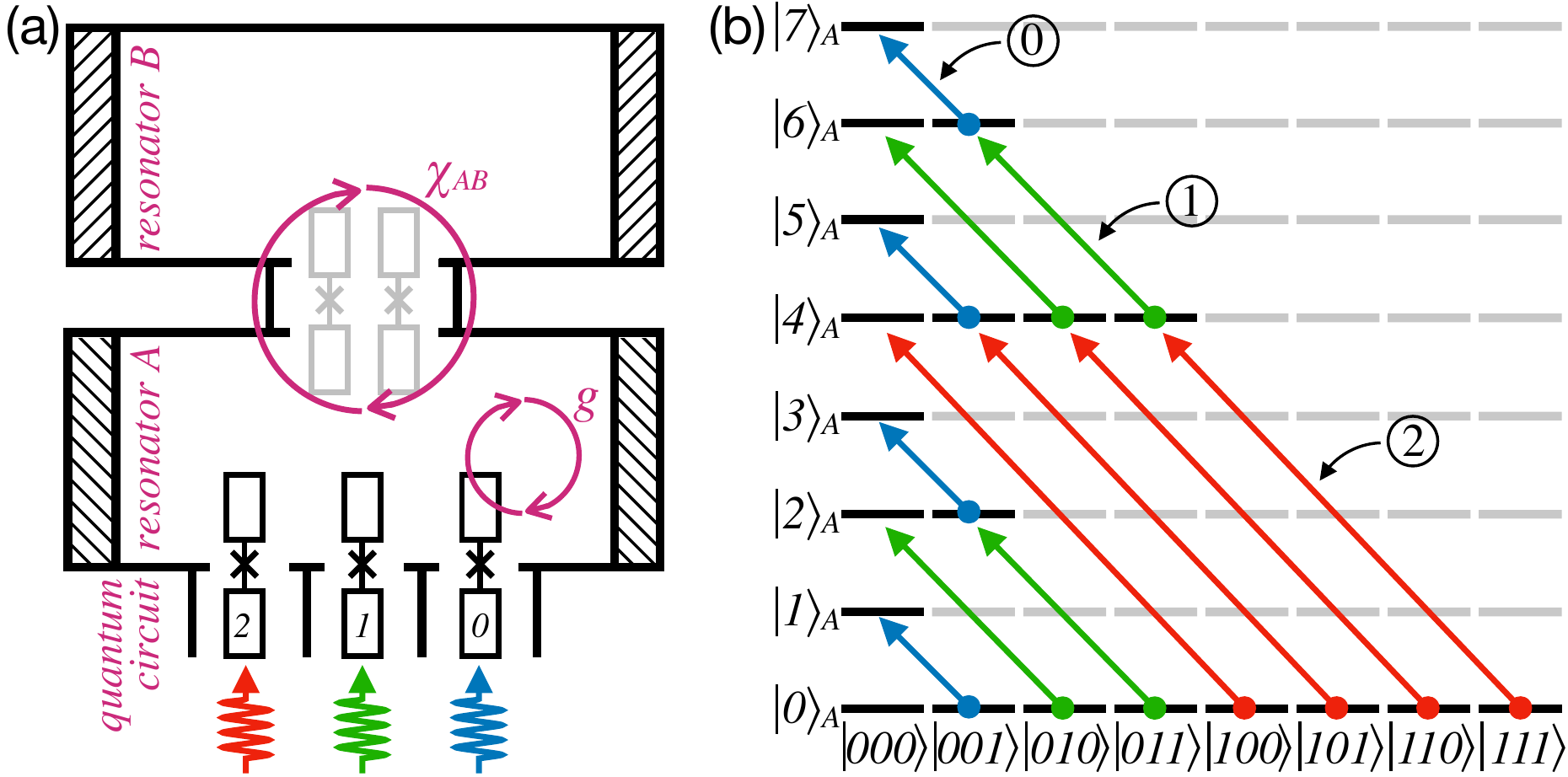}
  \caption{(a) Physical implementation of the QFT proposal in superconducting quantum circuits, where each qubit is coupled to the resonator A with coupling strength $g$. The two resonators A and B are coupled by a cross-Kerr interaction $\chi_{\rm AB}$, which is realized by mediating a $N$-type system, for example, two qubits, between them. (b) State transfer procedure in a $3$-qubit example. The horizontal and vertical axises label the multi-qubit states and the photon number states in A, respectively. The red/green/blue-colored arrows, labelled as \textcircled{\scriptsize 2}/\textcircled{\scriptsize 1}/\textcircled{\scriptsize 0}, indicate different control fields acting on the 2nd, 1st, and 0th qubit. The excitation of these qubits will be transferred into $4$, $2$, and $1$ photons in A, respectively.}
  \label{fig:schematic}
\end{figure}

Here, we propose a fully-quantum implementation of QFT for arbitrary $q$ by exploiting the infinite-dimensional Hilbert spaces of two coupled harmonic resonators. Figure\,\ref{fig:schematic}(a) shows a possible implementation of our proposal with superconducting quantum circuits, which consists of three major steps: (i) Transfer an arbitrary multi-qubit state to the resonator A, (ii) switch on a cross-Kerr interaction between A and B for a suitable time, and (iii) localize the resulting QFT state in B by applying a projective measurement to A. Afterwards, the result can be either transferred back to the quantum circuit for further processing or read out directly from B, depending on the specific applications of QFT. 

The rest of this paper is arranged as follows: In Sec.\,\ref{sec:state_transfer}, we describe a protocol that transfers an unknown multi-qubit state to one resonator, and \textit{vice versa}. Then, QFT can be performed in two-coupled oscillating modes as described in Sec.\,\ref{sec:qft}. Next, we take the phase estimation algorithm as an example of applications of our protocol for QFT, and compare the results with the qubit-based methods reported in the literature. Finally, we summarize our study and discuss possible improvements of our proposal in Sec.\,\ref{sec:conclusions}. The model of our system and detailed derivations of the state-transfer method can be found in Appendix\,\ref{app:model}.

\section{State transfer between multiple qubits and one resonator}\label{sec:state_transfer}
To perform QFT in oscillating modes, we first describe the transfer of an unknown $n$-qubit state to a single resonator, which is also an important problem on its own right. This state can be expressed in a basis, where each basis state is labeled with a binary string indicating the states of the individual qubits. In the string, starting from the right, the $k$th bit represents a decimal number $2^k$. Our goal is to produce a superposition state in the resonator, where each basis state of the $n$ qubits is replaced by a Fock state corresponding to the decimal number of the binary string. For example, in a $3$-qubit system the basis state $|110\rangle$ (binary string) is mapped to the Fock state $|6\rangle_{\rm A}$ (decimal number) in the resonator. The state mapping procedure is then composed of $n$ steps. Specifically, we couple the qubits on resonance with resonator A in \emph{reverse} order, i.e., $k=n-1,\cdots,0$. For each $k$, we transfer the population of the states $|m_k, 1\rangle_{\mathrm{A},k}$, where $m_k$ and $1$ denote the number of excitations in the resonator A and the $k$th qubit, to the new state $|m_k+2^k, 0\rangle_{\mathrm{A},k}$. This is equivalent to the binary-decimal number transform in mathematics, see Fig.\,\ref{fig:schematic}(b) for a $3$-qubit example.

\subsection{The all-resonant and perfect state transfer}
The details of our transfer protocol are described by one of the two following scenarios. For $k > 0$, we adiabatically tune the $k$th qubit into resonance with the resonator A while keeping the other qubits detuned, such that a suitable Jaynes-Cummings interaction can be generated between the two components. During this process, the initially uncoupled eigenstates  $|m_k,0\rangle_{\mathrm{A}, k}$ and $|m_k,1\rangle_{\mathrm{A},k}$ are transformed into the dressed states $|m_k,-\rangle_{\mathrm{A},k}$ and $|m_k+1,+\rangle_{\mathrm{A},k}$, respectively, where $m_k$ indicates the photon number in the resonator before coupling to the $k$th qubit. Here, we define the eigenstates of the resonant Jaynes-Cummings Hamiltonian as $|0,0\rangle_{\mathrm{A},k}$ for the ground state and $|m_k, \pm \rangle_{\mathrm{A},k} \equiv \left( |m_k,0\rangle_{\mathrm{A},k} \pm |m_k-1,1\rangle_{\mathrm{A},k} \right)/\sqrt{2}$ for $m_k>0$. The corresponding eigenenergies are $E_{m_k,\pm} = \hbar \left( m_k\omega_{\rm A} \pm \sqrt{m_k}g \right)$,  where $\omega_{\rm A}$ is the frequency of the resonator A and $g$ is the coupling rate. To transfer the population of this qubit to the resonator, we apply the following multi-frequency control to the qubit (see Appendix\,\ref{app:model} for derivations)
\begin{equation}
	V_{\rm map}^{(k)} = \sum_{m_{k}}\sum_{l=1}^{2^k-1}(-1)^{l-1} \hbar 2\Omega\sqrt{l(2^k-l)}\cos(\omega_{m_k,l} t) \sigma_{y,k}. \label{eq:all_resonant}
\end{equation}
Here, $\omega_{m_k,l} = (E_{m_{k}+l+1,+} - E_{m_{k}+l,-})/\hbar$ for even $l$ and $\omega_{m_k,l} = (E_{m_{k}+l+1,-} - E_{m_{k}+l,+})/\hbar$ for odd $l$, $\Omega$ is a scaling factor and $\sigma_{y,k}$ is the standard Pauli operator for the $k$th qubit. As shown in Fig.\,\ref{fig:schematic}(b), only resonator states $|m_{k}\rangle_{\rm A}$ with $m_k \in \{ 0, 2^{k+1}, \cdots, (2^{n-(k+1)}-1)2^{k+1} \}$ and $m_k \leq 2^{n-1}$ can have finite occupation before coupling the $k$th qubit. For each $m_k$, the control field is equivalent to the so-called \textit{all-resonant} control \cite{Strauch2012}, which couples all the states $|m_{k}+1,+\rangle, |m_{k}+2,-\rangle, \cdots, |m_{k}+2^k,-\rangle$ in a so-called \textit{perfect} state-transfer chain \cite{Christandl2004, *Christandl2005}. Note that by coupling the qubits in a reverse order, the values of $m_k$ are sufficiently separated from each other. Hence, each driving frequency $\omega_{m_k,l}$ in Eq.\,\eqref{eq:all_resonant} is unique and the state transitions for all $m_{k}$ can happen in parallel. Most importantly, by applying the control field $V_{\rm map}^{(k)}$ for a time period $\tau_{\rm map}=\pi/\Omega$, we can realize the transitions $|m_{k}+1,+\rangle_{\mathrm{A},k} \rightarrow |m_{k}+2^k,-\rangle_{\mathrm{A},k}$ for all $m_{k}$ within a \emph{single} step.  The $|m_{k},-\rangle_{\mathrm{A},k}$ states are not influenced because they are not in any of the transition chains. Finally, at the end of each step, we adiabatically detune the $k$th qubit from the resonator. As a consequence, the dressed states $|m_{k},-\rangle_{\mathrm{A},k}$ and $|m_{k}+2^k,-\rangle_{\mathrm{A},k}$ are transformed to the uncoupled states $|m_{k},0\rangle_{\mathrm{A},k}$ and $|m_{k}+2^k,0\rangle_{\mathrm{A},k}$. For our $3$-qubit example, the coefficient of $|110\rangle$ is mapped to the resonator state via the two steps: $|0\rangle_{\rm A}|110\rangle \rightarrow |4\rangle_{\rm A}|010\rangle\rightarrow |6\rangle_{\rm A}|000\rangle$. 

For $k=0$, the interaction described by Eq.\,\eqref{eq:all_resonant} vanishes because the desired transitions are photon number preserving. After adiabatically tuning the $0$th qubit into resonance with resonator A, we therefore apply the interaction
\begin{equation}
	V_{\rm map}^{(0)} = \sum_{m_0}\hbar 2\sqrt{2}\Omega 
	\left( -\cos(\omega_+ t) + \cos(\omega_1 t) \right) \sigma_{y,0}, \label{eq:photon_preserving}
\end{equation}
to the qubit. Physically, the frequency $\omega_{+}=(E_{m_{0}+2,+} - E_{m_{0}+1,+})/\hbar$ induces a jump in the Jaynes-Cummings ladder and the frequency $\omega_1=(E_{m_{0}+2,+} - E_{m_{0}+1,-})/\hbar$ corresponds to a qubit rotation. The drive $V_{\rm map}^{(0)}$ couples the three states $|m_0+1,+\rangle_{\mathrm{A},0}\rightarrow |m_0+2,+\rangle_{\mathrm{A},0} \rightarrow |m_0+1,-\rangle_{\mathrm{A},0}$ in a chain, where the last two states are empty before applying the driving field. Similar to the $k>0$ case, we detune the $0$th qubit after $\tau_{\rm map}$ to implement the transition $|m_{0}+1,-\rangle_{\mathrm{A},0} \rightarrow |m_{0}+1,0\rangle_{\mathrm{A},0}$. Then, all the qubits are off-resonant to the resonator A, inhibiting any further qubit-resonator interactions. All in all, the $n$-qubit state is successfully transferred to the resonator A by $n$ steps, while the qubits are automatically reset to the ground state.

\subsection{Error analyses}
The performance of our protocol can be benchmarked using the experimentally achievable parameters in typical superconducting quantum circuits. We choose $\Omega/2\pi \simeq 5\,\mathrm{MHz}$ and estimate the total driving time as $n\times 100\,\mathrm{ns}$ for an arbitrary $n$-qubit quantum state transfer. On the other hand, considering a qubit anharmonicity $\alpha/2\pi \simeq -200\,\mathrm{MHz}$ and a coupling strength $g/2\pi \simeq 200\,\mathrm{MHz}$, the adiabatic frequency tuning requires $\tau_{\rm ad} \simeq 100\,\mathrm{ns}$ to fulfill the adiabatic condition $1/\tau_{\rm ad} \ll |g|, |\alpha|$. In total, we estimate the time required for a $n$-qubit state transfer to $\tau_1 \simeq n\times 300\,\mathrm{ns}$. Hence, with the state-of-the-art coherence times of qubits exceeding $50\,\mathrm{\mu s}$ \cite{Axline2016} and single-photon lifetimes in 3D microwave resonators up to around $10\,\mathrm{ms}$ \cite{Reagor2013, *Reagor2016}, we estimate that a $10$-qubit state mapping could be achievable in current experiments, which requires qubit and single-photon lifetimes larger than $3\,\mathrm{\mu s}$ and $3\,\mathrm{ms}$, respectively. This results has taken the $1/m$ dependence for the lifetime of the $m$-photon Fock state into consideration.

Besides decay and decoherence of the quantum devices, experimental imperfections may also be considered for evaluating the transfer fidelity. Without an exact knowledge of how the different error sources may influence the results in real experiments, we consider the same error model studied in the perfect state transfer method \cite{Christandl2005} and focus on (i) the timing jitter of the driving field and (ii) the fluctuation of the energy levels. For the error source, (i), we recall that the state transfer fidelity at time $t$ in a $n'$-node chain is \cite{Christandl2004, *Christandl2005}
\begin{equation}
	F_{n'}(t) = \left[ \sin \left(\Omega t/2 \right) \right]^{2(n'-1)}.
\end{equation}
Considering a small timing jitter $\delta t$ and use Taylor series to expand the fidelity to the second order of $\delta t$ around $t_0=\pi/\Omega$, we have
\begin{equation}
	F_{n'}\left( t_0 +\delta t \right) \approx 1 - \frac{\left(n'-1\right)\pi^2}{4} \left(\frac{\delta t}{t_0}\right)^2. \label{eq:transfer_chain}
\end{equation}
Depending on the exact multi-qubit state to the transferred, from $0$ to $n-1$ number of parallel perfect state transfer paths can be constructed during the transfer process. For example, to transfer the $3$-qubit state $\left(|010\rangle + |101\rangle\right)/\sqrt{2}$, one effectively constructs two paths $|0\rangle_{\rm A}|010\rangle \rightarrow |2\rangle_{\rm A}|000\rangle$ and $|0\rangle_{\rm A}|101\rangle \rightarrow |4\rangle_{\rm A}|001\rangle \rightarrow |5\rangle_{\rm A}|000\rangle$. In each path, there may also exist $1$ to $n$ transfer chains. More specifically, the first transfer path in the $3$-qubit example contains $1$ transfer chain with $2$ nodes, and the other contains $2$ chains with node numbers being $4$ and $3$, respectively. In these regards, the total transfer fidelity in this example is $F=\left[ F_{2}^2 + \left(F_{3}F_{4}\right)^2 \right]/2$. Here, the subscript $3$ indicates the photon-preserving transfer, i.e., the $k=0$ case. In what follows, we define $F_1 \equiv F_3$ to simplify the notation.

Let us consider the most difficult task that involves all the possible transfer paths, where the qubits are initially prepared at $\left( |0\cdots 00\rangle + |0\cdots 01\rangle + \cdots + |1\cdots 11\rangle \right)/\sqrt{2^n}$. To transfer the state to the resonator, one should construct $n-1$ paths, while each of which contains $1$ to $n$ transfer chains with different number of nodes. The total transfer fidelity can be written as 
\begin{align}
	F\left( t_0 +\delta t \right) &= \frac{1}{2^n}\left[ 1 + F_1^2 + F_2^2 + \cdots + F_{2^{n-1}}^2 \right.\nonumber \\
	&\left. + \left(F_1F_2 \right)^2 
	+ \left(F_1F_4 \right)^2 + \cdots + \left(F_{2^{n-2}} F_{2^{n-1}} \right)^2 \right. \nonumber \\ 
	&\left. + \cdots + \left(F_1 F_2 \cdots F_{2^{n-1}} \right)^2 \right].
\end{align}
This formula can be approximated by recalling the relation in Eq.\,\eqref{eq:transfer_chain}, that is
\begin{equation}
	F\left( t_0 +\delta t \right) \approx 1 - \frac{\pi^2}{2^{n+1}} \frac{\left[ 2^{n} - (n-1)\right]\left(2^n-1\right)}{n}\left( \frac{\delta t}{t_0}\right)^2.
\end{equation}
Here, the approximation is valid to the fourth order accuracy. We observe that the leading-order error of the transfer fidelity is at the scale of $\delta t^2$, which is almost independent of the number of qubits, $n$. However, the coefficient of $\delta t^2$ may scale with the qubit number, $n$, as $2^{n-1}/n^3$ if we further limit the driving strength to be $t_0 \propto n$ \cite{Christandl2005}. In this regard, one may conclude that a small time jitter, $\delta t$, is crucial for applying the state-transfer method in increasingly larger-dimensional systems.

Next, we consider the error source, (ii). We assume a time-independent energy fluctuation $\delta E$ for all the energy levels of the system. To quantify this influence to the transfer fidelity, we let the system evolve for a time period of $2t_0$ and calculate the absolute square of the overlap between the initial and the final states \cite{Christandl2005}. In the ideal case in a $N$-node transfer chain, the initial state at the first node will be perfectly transferred to the final node at $t_0$, then transfered back to the initial node at $2t_0$ with fidelity $1$. However, with the fluctuations in the energy levels, the actual fidelity in a $n'$-node chain is \cite{Christandl2005}
\begin{equation}
	F_{n'}\left( \delta E \right) = 1 - 2t_0 \delta E.
\end{equation}
Considering the parallel transfer chains that may exist in our protocol, the fidelity function may be written in a slightly different form 
\begin{equation}
	F\left( \delta E \right) = \frac{\left[1 + \left( 1 - 2t_0 \delta E \right)^2\right]^n}{2^{n}} 
	\approx 1 - nt_0\delta E.
\end{equation}
Here, the approximation is valid to the second order accuracy. The leading-order error of the transfer fidelity scales linearly with both the energy fluctuation, $\delta E$, and the number of qubits, $n$. However, the fidelity scales with $n^2$ with the limit of the driving strength, $t_0 \propto n$.

\subsection{Example: Transfer of a three-qubit state}
\begin{figure}[h]
  \centering
  \includegraphics[width=\columnwidth]{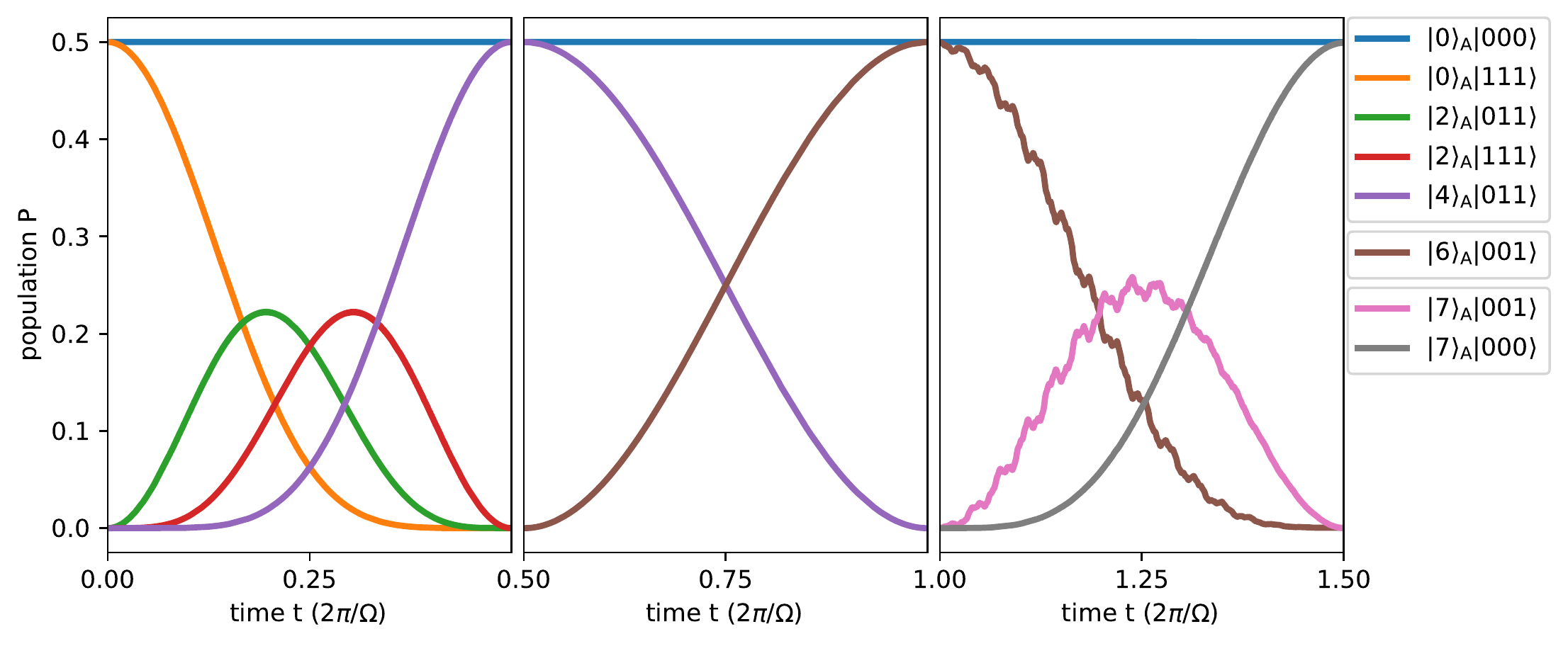}
  \caption{The state transfer process of a three-qubit example, where the qubits are initially prepared at the superposition state $\left(|000\rangle + |111\rangle\right)/\sqrt{2}$ and the resonator is at the vacuum state $|0\rangle_{\rm A}$. The transfer fidelities of the three steps, from $|0\rangle_{\rm A}\left(|000\rangle + |111\rangle\right)/\sqrt{2}$ to $\left(|0\rangle_{\rm A}|000\rangle + |4\rangle_{\rm A}|011\rangle\right)/\sqrt{2}$, $\left(|0\rangle_{\rm A}|000\rangle + |6\rangle_{\rm A}|001\rangle\right)/\sqrt{2}$, and $\left(|0\rangle_{\rm A} + |7\rangle_{\rm A}\right)|000\rangle/\sqrt{2}$, are $1.0000$, $1.0000$, and $0.9992$, respectively. We note that the states are labelled in the uncoupled basis of the resonator and the qubits for the simplicity of illustration. The horizontal axis is labelled in unit of $2\pi/\Omega$.}
  \label{fig:stf}
\end{figure}

For clarity, we describe the transfer process of a three-qubit state as an example. Initially, the quantum information are encoded in the arbitrary $3$-qubit state
\begin{equation}
	|\psi(0)\rangle = \sum_{m=0}^{7} c_m |m\rangle, \label{eq:initial_state}
\end{equation}
where $m$ can be written in either decimal or in binary units. To transfer $|\psi\rangle$ to the resonator A, we consider first a subsystem composed of the $2$nd qubit and the resonator. The other qubits are largely detuned from the subsystem such that their states keep invariant in the interaction picture. When tuning the qubit and the resonator on resonance, the initial state of the whole system can be written as
{\small \begin{align}
	|\psi(0)\rangle = &|1,+\rangle_{\mathrm{A},2}
	\left( c_7|3\rangle_{1,0} + c_6|2\rangle_{1,0} 
	+ c_5|1\rangle_{1,0} + c_4|0\rangle_{1,0}  \right) 
	\nonumber \\
	+ &|0,-\rangle_{\mathrm{A},2}
	\left( c_3|3\rangle_{1,0} + c_2|2\rangle_{1,0} 
	+ c_1|1\rangle_{1,0} + c_0|0\rangle_{1,0}  \right),
\end{align}}%
where $|0\rangle_{1,0}=|00\rangle$, $|1\rangle_{1,0}=|01\rangle$, $|2\rangle_{1,0}=|10\rangle$, $|3\rangle_{1,0}=|11\rangle$ with the subscript indicating the corresponding qubit(s) number. We apply the multi-frequency control pulses described in Eq.\,\eqref{eq:all_resonant} to transfer $|1,+\rangle_{\mathrm{A},2}$ to $|4,-\rangle_{\mathrm{A},2}$ while leaving $|0,-\rangle_{\mathrm{A},2}$ unchanged. Next, we adiabatically detune them in a way that the dressed states $|4,-\rangle_{\mathrm{A},2}$ and $|0,-\rangle_{\mathrm{A},2}$ become $|4\rangle_{\rm A}|0\rangle_{2}$ and $|0\rangle_{\rm A}|0\rangle_{2}$, respectively. The state of the whole system thus reads
{\small \begin{align}
	|\psi(\tau_{1})\rangle = 
	\big[ &|4\rangle_{\rm A}
	\left( c_7|3\rangle_{1,0} + c_6|2\rangle_{1,0} 
	+ c_5|1\rangle_{1,0} + c_4|0\rangle_{1,0}  \right) 
	\nonumber \\
	+ &|0\rangle_{\rm A}
	\left( c_3|3\rangle_{1,0} + c_2|2\rangle_{1,0} 
	+ c_1|1\rangle_{1,0} + c_0|0\rangle_{1,0} \right) \big]|0\rangle_{2}.
\end{align}}%

In the second step, we adiabatically tune the $1$st qubit into resonance with the resonator, and obtain
\begin{align}
	|\psi(\tau_{1})\rangle = 
	\big[ &|5,+\rangle_{\mathrm{A},1}
	\left( c_7|1\rangle_{0} + c_6|0\rangle_{0} \right)
	\nonumber \\
	+ &|4,-\rangle_{\mathrm{A},1}
	\left( c_5|1\rangle_{0} + c_4|0\rangle_{0}  \right)
	\nonumber \\
	+ &|1,+\rangle_{\mathrm{A},1}
	\left( c_3|1\rangle_{0} + c_2|0\rangle_{0} \right)
	\nonumber \\
	+ &|0,-\rangle_{\mathrm{A},1}
	\left( c_1|1\rangle_{0} + c_0|0\rangle_{0}  \right) \big]|0\rangle_{2}.
\end{align}
By using the same method, we realize the state transitions $|5,+\rangle_{\mathrm{A},1} \rightarrow |6,-\rangle_{\mathrm{A},1}$ and $|1,+\rangle_{\mathrm{A},1} \rightarrow |2,-\rangle_{\mathrm{A},1}$. After the adiabatic detuning procedure, we obtain 
\begin{align}
	|\psi(2\tau_{1})\rangle = 
	\big[ &|6\rangle_{\rm A}
	\big( c_7|1\rangle_{0} + c_6|0\rangle_{0} \big)
	\nonumber \\
	+ &|4\rangle_{\rm A}
	\big( c_5|1\rangle_{0} + c_4|0\rangle_{0}  \big)
	\nonumber \\
	+ &|2\rangle_{\rm A}
	\big( c_3|1\rangle_{0} + c_2|0\rangle_{0} \big)
	\nonumber \\
	+ &|0\rangle_{\rm A}
	\big( c_1|1\rangle_{0} + c_0|0\rangle_{0}  \big) \big]|0\rangle_{2,1}. \label{eq:1st_qubit}
\end{align}

In the last step, we wish to transform the excitation of the $0$th qubit to a single extra photon in the resonator A, which is photon preserving and cannot achievable by using the ``all-resonant" control method. However, the basic idea remains that one should construct $2^{n-1}$ parallel state-transfer chains between the initial and the final state, and use the perfect state transfer condition to determine the optimal coupling strength between each pair of nodes. One can verify that the multi-frequency control field described by Eq.\,\eqref{eq:photon_preserving} couples the three states $|m_0+1,+\rangle_{\mathrm{A},0}\rightarrow |m_0+2,+\rangle_{\mathrm{A},0} \rightarrow |m_0+1,-\rangle_{\mathrm{A},0}$ in a perfect state transfer chain, where the last two states are empty before applying the field, $m_0=0,2, 4, 6$ is the possible photon numbers in the resonator that is the result of the first two steps. Thus, by adiabatically tuning the resonator A and the $0$th qubit into resonance and applying the above field to the $0$th qubit, we realize the state transition $|m_0+1,+\rangle_{\mathrm{A},0}\rightarrow |m_0+1,-\rangle_{\mathrm{A},0}$. The final state after the adiabatic detuning is
\begin{equation}
	|\psi(3\tau_{1})\rangle = 
	\left( \sum_{m=0}^{7}c_m|m\rangle_{\rm A} \right) 
	|0\rangle_{2,1,0}. \label{eq:final_state}
\end{equation}
By comparing with the initial $3$-qubit state shown in Eq.~\eqref{eq:initial_state}, we see that an arbitrary $3$-qubit state is perfectly transferred to the resonator after $3$ steps.

In Fig.\,\ref{fig:stf}(a)-(c), we numerically simulate the $3$-step state transfer process in the $3$-qubit example with all the parameters, except the driving strength $\Omega$, the same with that in the error analysis. Here, the initial $3$-qubit state $\left(|000\rangle + |111\rangle\right)/\sqrt{2}$ is transferred to the resonator state $\left(|0\rangle_{\rm A} + |7\rangle_{\rm A}\right)/\sqrt{2}$, through the intermediate resonator-qubits states $\left(|0\rangle_{\rm A}|000\rangle + |4\rangle_{\rm A}|011\rangle\right)/\sqrt{2}$ and $\left(|0\rangle_{\rm A}|000\rangle + |6\rangle_{\rm A}|001\rangle\right)/\sqrt{2}$. Our simulation is based on the effective Hamiltonian Eq.\,\eqref{eq:hamiltonian_without_rwa} without rotating wave approximation, which explains the small ripples in the state populations during the transfer process. The ripples are evident in the $3$rd step, for that the transition frequencies $\omega_{+}$ is less distinguishable than $\omega_{l}$ for large $m_0$. In this case, the selectivity of the driving pulses decreases and the influence of the counter rotating terms becomes evident. This is the reason that we choose a smaller driving strength, $\Omega=200\,{\rm kHz}$, in the simulation than the value that has been used for error analysis. However, we note that $\Omega=5\,{\rm MHz}$ still applies to the first two steps, which leads to fidelities of $0.9973$ and $0.9993$, respectively. A larger $\Omega$ for faster state transfer process requires one to employ the optimal-control method, as illustrated in Refs.~\onlinecite{Strauch2012, Motzoi2013}, use different $\Omega$ in different steps and tune their values slowly with time so as to improve the frequency selectivity. A more quantitive appreciation of how optimal control improves the transfer fidelity should depend on detailed circuit parameters and possible constraints and lies outside the major interests of this paper.

\section{QFT in coupled resonators}\label{sec:qft}
\subsection{Performing QFT with cross-Kerr interaction}
Upon transfer of an $n$-qubit state using the method described above, the state of resonator A is
\begin{equation}
	|\psi(\tau_1)\rangle_{\rm A} = \sum_{m=0}^{q-1} c_m |m\rangle_{\rm A}. 
\end{equation}
By definition, the QFT acting on $|\psi(\tau_1)\rangle_{\rm A}$ should result into the state $\sum_{m=0}^{q-1} c_m |\mathcal{F}(m)\rangle$, where 
\begin{equation}
	| \mathcal{F}(m) \rangle = \left( \frac{1}{\sqrt{q}}\sum_{n=0}^{q-1} e^{i2\pi m n/q} |n\rangle\langle m| \right)|m\rangle. \label{eq:fourier_transform}
\end{equation}

We describe the coupling of a second resonator B to A by the cross-Kerr interaction 
\begin{equation}
	V_{\rm Kerr} = \hbar \chi_{\rm AB} a^{\dag}a b^{\dag}b, 
	\label{eq:kerr}
\end{equation}
where $\chi_{\rm AB}$ is the coupling strength between the two resonators. In superconducting quantum circuits, a cross-Kerr interaction with strength $\chi_\mathrm{AB}/2\pi \simeq 2.5\,\mathrm{MHz}$ has been reported by coupling $N$-type system to both resonators, e.g., two qubits as shown in Fig.\,\ref{fig:schematic}(a) \cite{Rebifmmodecuteclseci2009, Hu2011}. However, as will be discussed later, a weak cross-Kerr interaction around several kilohertz is already sufficient for implementing QFT in our protocol. We assume that the resonator B is prepared in the state 
\begin{equation}
	|\psi(\tau_1)\rangle_{\rm B}= \frac{1}{\sqrt{q}} (|0\rangle_{\rm B} + |1\rangle_{\rm B} + \cdots + |q-1\rangle_{\rm B}), \label{eq:initial_b}
\end{equation}
which may be realized by employing one of the methods reported in Refs.\,\onlinecite{Liu2004, Hofheinz2008, *Hofheinz2009, Strauch2010, *Strauch2012a, *Sharma2016, Wang2017}. The free evolution of the two-mode system in the interaction picture reads
\begin{equation}
	|\psi(\tau_1+t)\rangle_{\rm AB} = \sum_{m=0}^{q-1} c_m |m\rangle_{\rm A} 
	\frac{1}{\sqrt{q}} \sum_{n=0}^{q-1} e^{-i\chi_{\rm AB} t m n}|n\rangle_{\rm B}.
\end{equation}
After waiting for a time period $\tau_2 = \left(-2\pi/q + 2k\pi \right)/\chi_{\rm AB}$ with $k$ being an arbitrary  integer such that $\tau_2>0$, we obtain the state 
\begin{equation}
	|\psi(\tau_1+\tau_2)\rangle_{\rm AB} = \sum_{m=0}^{q-1} c_m |m\rangle_{\rm A} |\mathcal{F}(m)\rangle_{\rm B}. \label{eq:entangled}
\end{equation}
Then, we control the $N$-type system between the two resonators to suppress the cross-Kerr interaction as described in Refs.\,\onlinecite{Rebifmmodecuteclseci2009, Hu2011}. For each $m$, one can immediately see that the states in resonator B correspond to the Fourier transform of the basis states $|m\rangle_{\rm A}$ as defined in Eq.\,\eqref{eq:fourier_transform}. Similarly, the inverse quantum Fourier transform ($\rm QFT^{-1}$) can be realized by following the same process but setting $\tau_2 = \left(+2\pi/q+2k\pi\right)/\chi_{\rm AB}$.

If we assume $\chi_{\rm AB}/2\pi=\mp 50\,\mathrm{kHz}$ with the sign controlled by the $N$-type system, the QFT or $\rm QFT^{-1}$ processes require a time duration of $\tau_2 \simeq (20/q)\,\mathrm{\mu s}$. Considering the $1/m$ dependence for the lifetime of the $m$-photon Fock state, our method requires a single-photon lifetime larger than $20\,\mathrm{\mu s}$, independent of the dimension of the Hilbert space $q$. Taking into account the higher-order nonlinearities that may exist in the resonator and the potential use of optimal control methods, we expect the required single-photon lifetime not to exceed $\sim 100\,\mathrm{\mu s}$ \cite{Warren1993, *Rabitz2000, Heeres2017}. Note that this value is still one order of magnitude smaller than the value of about $3\,\mathrm{ms}$ estimated above for the transfer of a $10$-qubit state. Hence, we conclude that it is the state transfer process that limits the achievable dimension $q$. This indicates that only a relatively weak cross-Kerr interaction is required in the system, which leaves substantial freedom in sample design and fabrication.

\subsection{Disentangling the two resonators}
To further extract the QFT state from the two resonators, one needs to disentangle the two resonators while keeping the coherence among different $|\mathcal{F}(m)\rangle_{\rm B}$. One possible solution may be projecting the resonator A onto the state $|p\rangle_{\rm A} = \sum_{m=0}^{q-1}|m\rangle_{\rm A}/\sqrt{q}$, which is realized by two steps. In the first step, we reverse the state transfer method introduced above to transfer the entanglement between the resonators A and B to that between the $n$ qubits in the quantum circuit and resonator B. This procedure requires a time duration of $\tau_1$. Next, we apply $n$ projective measurements along the X-direction for each qubit, which effectively realizes the projection $|p\rangle\langle p|$ on the $n$ qubits. The detection of the qubits in state $|p\rangle$ results in the following state of the whole system 
\begin{equation}
	|\psi(2\tau_1+\tau_2+\tau_3)\rangle = |0\rangle_{\rm A} \otimes \left(\sum_{m=0}^{q-1} c_m |\mathcal{F}(m)\rangle_{\rm B} \right) \otimes |p\rangle,
\end{equation}
where $\tau_3$ indicates the time duration for $n$ projective measurements on the qubits. In this way, the QFT state is entirely localized in resonator B. The success probability of the detection result $|p\rangle$ is $P=1/q$, which is independent of the specific state in the two resonators. 

For large $q$, this small probability implies a considerable number of repeated measurements. In each repeat, we reset the qubits and resonators, do state transfer, free evolution, and a set of projective measurements on qubits. Then, post-selection is needed to distinguish the successful measurements from the data. However, we note that the number of repeated measurements is not limited by the decoherence time of the system. Our method provides a possible way to realize high-dimensional and fully-quantum QFT at the price of a large number of repeated measurements, which is otherwise hardly achievable by using the existing methods because of the currently limited qubit lifetime. 

Alternatively, one may also employ the method of weak measurement \cite{Hatridge2013} with unitary qubit control to align all the qubits in almost the same direction, for example, the ground state. This procedure disentangles the resonator-B from the other circuit components with a larger success probability ideally approaching the unity. However, the finite lifetime of quantum information in our system may set a heavy constraint on the possible applications of this method. In these regards, we may leave it as an open question that whether there exists a simple but deterministic method to disentangle the two resonators without destroying the coherence of resonator B state.

\subsection{Example: QFT of an $3$-qubit state}
\begin{figure}[h]
  \centering
  \includegraphics[width=\columnwidth]{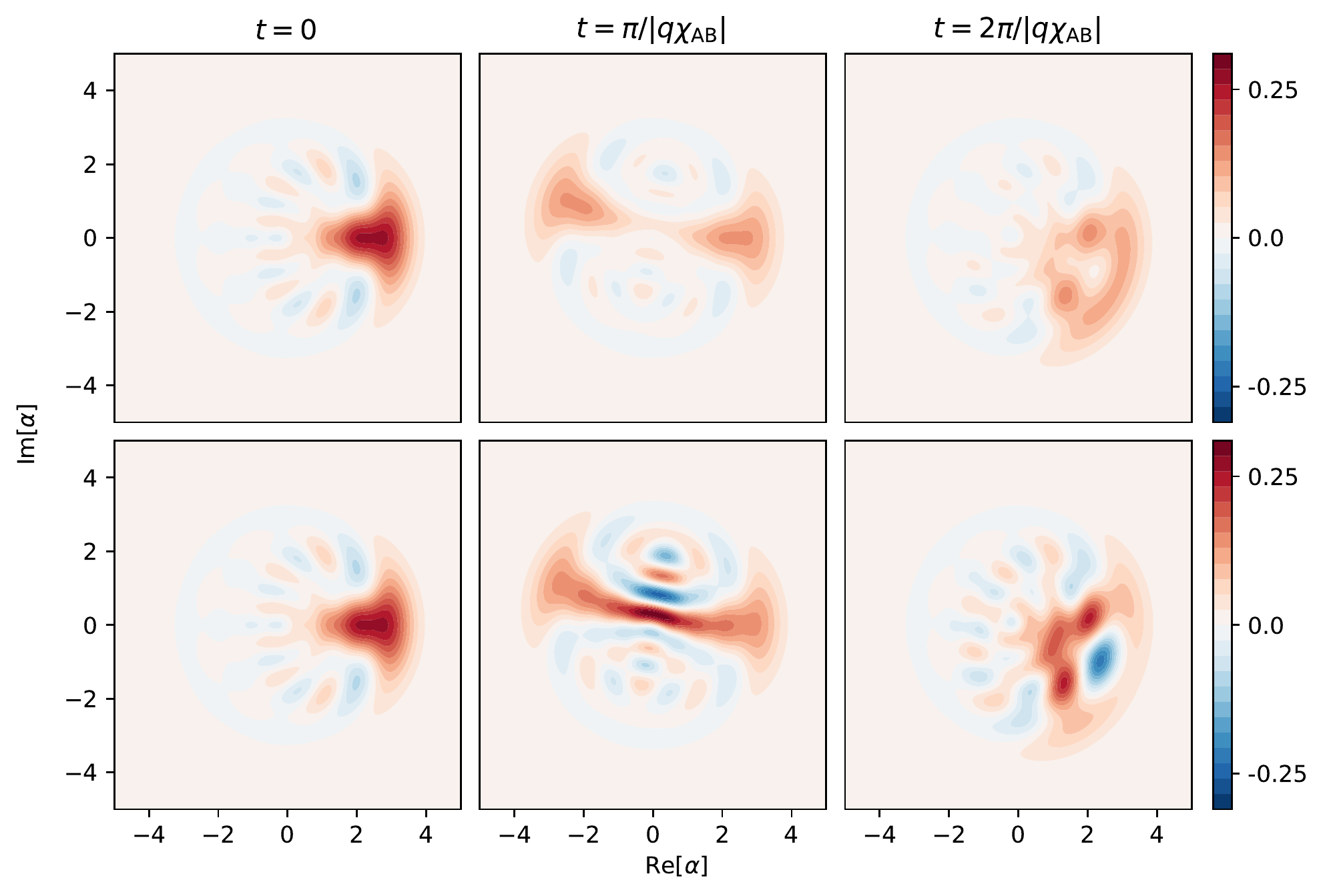}
  \caption{Time evolution of the resonator-B state for a $2^3$-dimensional QFT, where the two resonators are initially prepared at the states $\left(|000\rangle_{\rm A} + |111\rangle_{\rm A}\right)/\sqrt{2}$ and $(|0\rangle_{\rm B} + |1\rangle_{\rm B} + \cdots + |7\rangle_{\rm B})/\sqrt{8}$, respectively. Here, the top panel shows Wigner function of the partial trace of the two-resonator state over the resonator A. The bottom panel shows the QFT state at different time instances, where the two resonators are decoupled by a successful projective measurement.}
  \label{fig:qft}
\end{figure}

In Fig.\,\ref{fig:qft}, we simulate the time evolution of composite system for the $3$-qubit example with all the parameters the same as before. When taking a partial trace of the two-resonator state over the resonator A, we obtain a mixed states of the QFT states for each photon number in A. Specifically, at the time instance $t=-2\pi/\left(q\chi_{\rm AB}\right)$, we obtain a statistical mixture of the QFT states of all the possible photon numbers, i.e., $\sum_{m=0}^{q-1} \left|c_m\right|^2 |\mathcal{F}(m)\rangle_{\rm B}$. By comparison, the exact QFT state is obtained by performing a successful projective measurement on resonator A, that is, $\sum_{m=0}^{q-1} c_m |\mathcal{F}(m)\rangle_{\rm B}$.

\section{Using oscillating modes in QFT-related quantum algorithms}\label{sec:phase_estimation}
With the capability of performing QFT in oscillating modes, it is straightforward to apply this method in various quantum algorithms, such as quantum phase estimation algorithm, Shor's factorizing algorithm, and the HHL linear-equation solving algorithm. Here, we take the quantum phase estimation algorithm for illustration. Assuming that the unitary operator $U$ acting on the multi-qubit state $|\psi\rangle$ results in an unknown phase $\theta$, i.e., $U|\psi\rangle=e^{i\theta}|\psi\rangle$, the aim of quantum phase estimation is to determine $\theta$ by using $\rm QFT^{-1}$. To transfer the phase information from the quantum circuit to a single resonator, we build a sequence of controlled-$\rm U$ gates between the ancilla qubit and the quantum circuit. Before every controlled-$\rm U$ gate, the ancilla qubit is prepared in $|\psi\rangle_{\mathrm a}=\frac{1}{\sqrt{2}}\left(|0\rangle_{\mathrm a} + |1\rangle_{\mathrm a} \right)$ by a Hadamard gate. After each controlled-$\rm U$ gate, we employ the state mapping method and implement the following transition $|1\rangle_{\mathrm a} \otimes |m\rangle_{\rm A} \rightarrow |0\rangle_{\mathrm a}\otimes |m+2^k\rangle_{\rm A}$. This state mapping process also resets the ancilla qubit to the ground state, so that it can be recycled straightforwardly. After $n$ steps, we obtain the following state in the resonator A
\begin{equation}
	|\psi(\tau_1)\rangle_{\rm A}=\frac{1}{\sqrt{q}}\sum_{m=0}^{q-1} e^{im \theta} |m\rangle_{\rm A}. 
\end{equation}
Next, we couple the resonator B to A through a cross-Kerr interaction and implement $\rm QFT^{-1}$ in the oscillating modes. As discussed before, the free evolution for a time period $\tau_2$ takes the system to the following state
\begin{equation}
	|\psi(\tau_1+\tau_2)\rangle_{\rm AB} = \frac{1}{q} \sum_{m=0}^{q-1} |m\rangle_{\rm A} \sum_{n=0}^{q-1} 
	e^{i  m(\theta - 2\pi n/q)} |n\rangle_{\rm B}. 
\end{equation}
Finally, we perform a projective measurement on the resonator A to disentangle the two resonators and obtain the following state in B
\begin{equation}
	|\psi(2\tau_1+\tau_2+\tau_3)\rangle_{\rm B} \approx | \left[ q\theta/2\pi \right] \rangle_{\rm B},
\end{equation}
where $[q\theta/2\pi]$ represents the nearest integer to $q\theta/2\pi$. We note that this approximation becomes an equality when $q\theta/2\pi$ is an integer. This observation indicates that the phase $\theta$ can be estimated by measuring the photon number in resonator B. The estimation precision is $\pm \pi/q$, which is proportional to the inverse of the Hilbert space dimension $q$. 

\begin{figure}[t]
  \centering
  \includegraphics[width=\columnwidth]{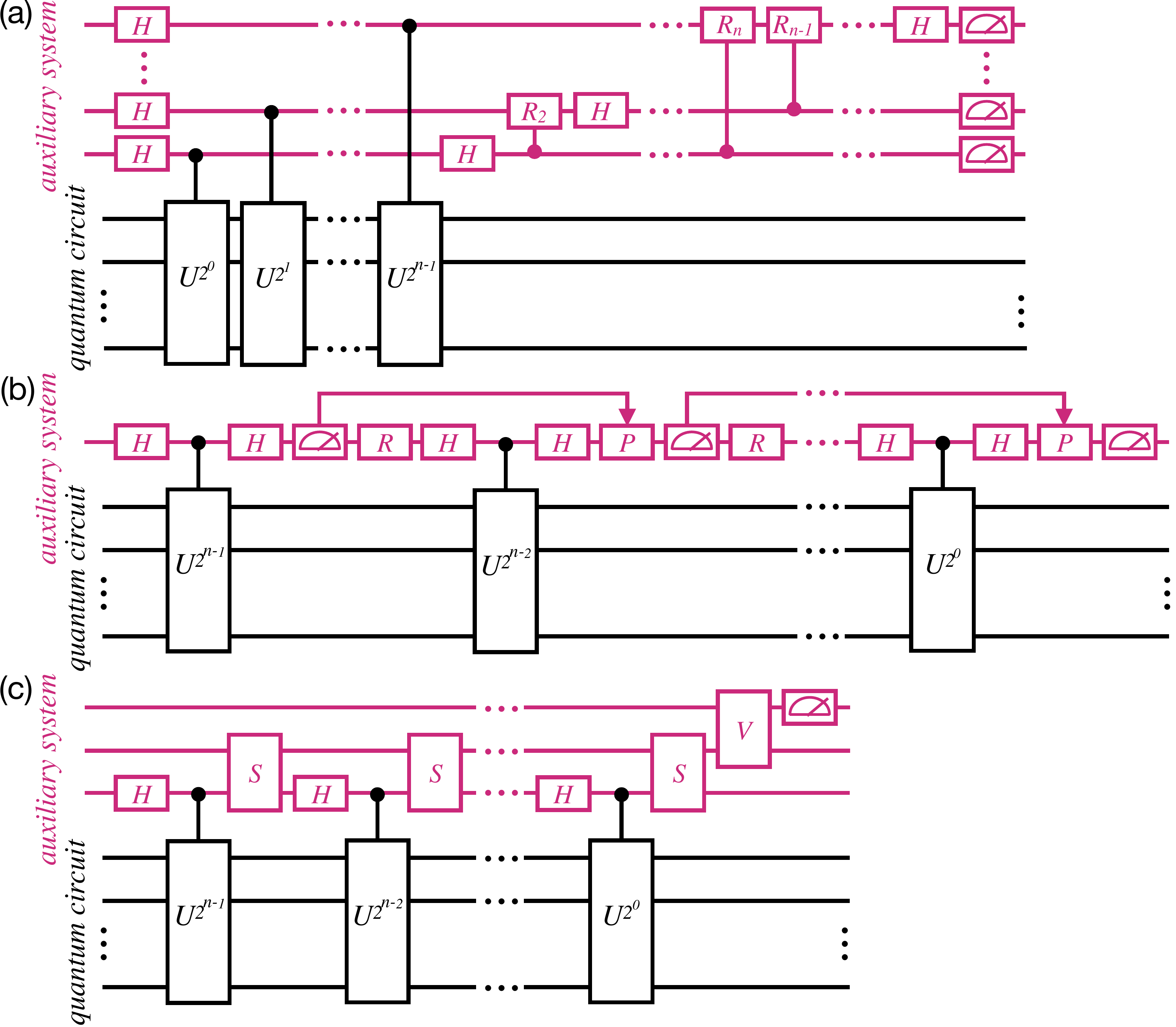}
  \caption{Circuit diagrams for the quantum phase estimation, using (a) the conventional approach, (b) qubit recycling and (c) our approach. Here, $H$ denotes the Hadamard gate, $U^{2^k}$ the controlled phase gate, $R_k$ the two-qubit gate, $R$ the qubit resetting gate, and $P$ the measurement-based single-qubit gate. The symbol $S$ denotes the state transfer process from a single qubit to the resonator A, and $V$ denotes the projective measurement on B, which is counted as $n$ single-qubit state transfer and $n$ projective measurements on the qubits.}
  \label{fig:cpr}
\end{figure}

In Fig.\,\ref{fig:cpr}, we compare the circuit diagrams for (a) the conventional approach, (b) qubit recycling, and (c) our approach in quantum phase estimation. For a Hilbert space with dimension $q=2^n$, the three approches require $n$ ancilla qubits, $1$ ancilla qubit, and $1$ ancilla qubit and $2$ resonators, respectively. Within the coherence time of the whole system, the conventional approach requires $2n$ Hadamard gates, $n(n-1)/2$ two-qubit gates for $\rm QFT^{-1}$ and $n$ measurements \cite{Nielsen2010}. Qubit recycling requires $2n$ Hadamard gates, $n$ real-time measurements, $(n-1)$ measurement-based single qubit gates, and $(n-1)$ qubit resetting gates \cite{Martin-Lopez2012}. Our approach requires $n$ Hadamard gates, $2n$ times of single-qubit state transfer, $n$ measurements on the qubits, and $1$ final measurement on the photon number in the ancilla resonator. The number of controlled-$\rm U$ gates is $n$, which is the same for the three approaches and should be added in the total number of operations. They are $n(n+7)/2$, $(6n-2)$, and $(5n+1)$, respectively. We note that these numbers may vary in specific realizations but the order of magnitude should remain the same.

The success probability in our approach is not unitary, which indicates that it should be repeated several times and the data post-selection is required to distinguish the successful measurements from the data. However, this repeat number is not limited by the coherence time, such that it is not counted in the comparison with the aim of realizing a high-dimensional QFT for high-precision phase estimation. In summary, the resources required by the three approaches are: (a) conventional approach: $n=\log_2(q)$ ancilla qubits and $n(n+7)/2$ operations, (b) qubit recycling: $1$ ancilla qubit and $(6n-2)$ operations, and (c) our approach: $1$ ancilla qubit and $2$ resonators, and $(5n+1)$ operations. We see that both (b) and (c) reduce the qubit resources and the number of operations to a large extent, as compared to (a). However, qubit recycling, (b), requires $n$ real-time measurements and measurement-based single-qubit gates within the coherence time of the quantum circuit, which is a significant challenge considering current fabrication, measurement, and control technologies. More importantly, it is indeed a semi-classical QFT method which cannot be directly used in the quantum algorithms where fully-quantum QFT is desired. The above observations reveal that our method, (c), not only saves hardware resources such as ancilla qubits and gates, but also realizes full-quantum QFT which is desired in many quantum algorithms. 

\textit{Note added.---} One may also consider employing the approximation techniques to implement QFT in qubit systems, called AQFT \cite{Barenco1996}, which reduces the gate number of (a) to $(2n-m)(m-1)/2$ at an unavoidable approximation error of $\left|\epsilon\right|\leq n/2^m$. This method may lead to an even more favorable scalability in certain scenarios considering the fact that most of the physical implementations of quantum algorithms are not exact. However, the study of exact realizations of QFT is important on its own right, and we can hardly compare our method with AQFT because they do not have the same error model.

\section{Conclusions and outlook.}\label{sec:conclusions}
In this work, we propose to use two harmonic resonators with a tunable cross-Kerr interaction to realize high-dimensional and fully-quantum QFT in a superconducting quantum circuit architecture. To be compatible with the qubit-based quantum circuit, we also propose a method to transfer an unknown multi-qubit states between the two components. The whole procedure consists of the multi-qubit state transfer to a single resonator, the free evolution in two coupled resonators, and a projective measurement in one resonator. Our protocol avoids the need for a large number of qubits, gates, and feedforward measurements as required in conventional QFT-related quantum algorithms, and realizes fully-quantum QFT in contrast to qubit recycling. Given the control parameters we choose and the $10\,\mathrm{ms}$ single-photon lifetimes in the microwave resonator reported in the literature, we estimate that a $2^{10}$-dimensional fully-quantum QFT can be reached with current state-of-the-art superconducting quantum circuits. This dimension estimate is mainly limited by the adiabatic tuning condition during the state-transfer process. Here, the qubit lifetimes are not a major concern, since coherence times of several microseconds are sufficient. In order to accommodate more qubits, one needs either longer resonator lifetimes or shorter state-transfer times. Whereas the former is beyond the scope of this discussion, the latter may be achieved by optimizing the circuit parameters or using the short-time techniques to accelerate the adiabatic frequency tuning process \cite{GueryOdelin2019}.

In general, a harmonic resonator has the advantage of a large dimension of the Hilbert space and a long coherence time. Our study of QFT demonstrates that, by properly combining it with other quantum circuits, it is possible to implement complex quantum operations regardless of the quantum circuit depth and limited qubit coherence times. This combination opens multitudes of new possibilities for realizing various kinds of useful quantum algorithms in the noisy intermediate-scale quantum (NISQ) devices.

\section*{Acknowledgements}
We acknowledge support by the German Research Foundation via Germany's Excellence Strategy EXC-2111-390814868, the Elite Network of Bavaria through the program ExQM, the European Union via the Quantum Flagship project QMiCS (Grant No.\,820505). R.B.W. acknowledges supports from the National Key Research and Development Program of China through Grant No.\,2017YFA0304304, and NSFC Grants No.\,61773232 and No.\,61833010. L.S. acknowledges the support from National Key Research and Development Program of China Grant No.\,2017YFA0304303 and National Natural Science Foundation of China Grant No.\,11474177. Y.X.L. is supported by Key-Area Research and Development Program of GuangDong Province under Grant No.\,2018B030326001. 

\appendix
\begin{widetext}
\section{The ``All-Resonant" and ``Perfect" State-Transfer Method}\label{app:model}
As it is schematically shown in the Fig.~1(a) in the main text, we consider a system with $n$ superconducting qubits as well as two microwave resonators coupled by a cross-Kerr interaction. The total Hamiltonian of such a system $H_{\rm S}=H_{0}+V$ can be written as
\begin{eqnarray}
	H_{0} &=& \hbar \omega_{\rm A}a^{\dagger}a + \hbar \omega_{\rm B}b^{\dagger}b + \frac{1}{2}\sum_{k=0}^{n-1}\hbar \omega_k\sigma_{z,k}, \\
	V &=& \sum_{k=0}^{n-1} \hbar g_k\left( a^{\dagger}\sigma_{-,k} + a\sigma_{+,k} \right) 
	+ \hbar \chi_{\rm AB} a^{\dagger}ab^{\dagger}b
\end{eqnarray} 
Here, $\omega_{\rm A}$ ($\omega_{\rm B}$), $a$ ($b$), and $a^{\dagger}$ ($b^\dagger$) are the resonance frequency, annihilation, and creation operators of the oscillating mode in resonator A (B), respectively. The symbols $\omega_k$, $\sigma_{z,k}$, $\sigma_{\pm,k}$ are the frequency and standard Pauli operators for the $k$th qubit in the ensemble. Furthermore, $g_k$ is the coupling rate between resonator A and the $k$th qubit and $\chi_{\rm AB}$ is the cross-Kerr interacting rate between the two resonators. We note that $g_k$ can be neglected in the rotating-wave approximation if the resonator A and the $k$th qubit are far detuned. The Kerr interaction can be turned on and off by controlling the $N$-type system coupled between the two resonators, with the method shown in Refs.\,\onlinecite{Rebifmmodecuteclseci2009, Hu2011}. Throughout this study, we use the interaction picture where $H_{\rm I} = e^{\frac{i}{\hbar} H_0 t} V e^{-\frac{i}{\hbar}H_0 t}$. 

The method of transferring an $n$-qubit state to the resonator A consists of $n$ steps. Here, we discuss the $(n-k)$th step where the excitation of the $k$th qubit is transformed into $2^k$ photons in the resonator. Let us suppose that the initial state of the $k$th qubit is $c_0|0\rangle_{k}+c_1|1\rangle_{k}$, where $c_0$ and $c_1$ are normalized amplitudes. By adiabatically tuning the qubit and the resonator on resonance, the total state of the combined system reads
\begin{equation}
	|\psi(0)\rangle_{\mathrm{A},k} 
	= \sum_{m} \left( c_0 |m,- \rangle_{\mathrm{A},k} + c_1 |m+1,+\rangle_{\mathrm{A},k} \right), 
\end{equation}
where $m=0, 2^{k+1}, \cdots, (n-1-k)2^{k+1}$ represents all the possible photon numbers that are already present in the resonator A before coupling the $k$th qubit. To proceed, we define the eigenenergies and eigenstates of the on-resonant Jaynes-Cummings Hamiltonian as
\begin{equation}
    E_{m,\pm} = \hbar \left( m\omega_{a} \pm \sqrt{m}g \right),~
	|m, \pm \rangle_{\mathrm{A},k} = \frac{1}{\sqrt{2}}\left( |m,0\rangle_{\mathrm{A},k} \pm |m-1,1\rangle_{\mathrm{A},k} \right),
\end{equation}
where $m$ and the number $0/1$ represent the excitation number in the resonator A and the $k$th qubit, respectively, and $g=g_k$ for the simplicity of notation.

As illustrated in the main text, for $k\neq 0$ we apply the following control field to the system
\begin{equation}
	V_{\rm map}^{(k)} = \hbar f_k(t) \sigma_{y,k},~
	f_k(t) = \sum_{m}\sum_{l=1}^{2^k-1} (-1)^{l-1} 2\Omega\sqrt{l(2^k-l)}\cos(\omega_l t)
\end{equation}
with $\omega_l = E_{m+l+1,+} - E_{m+l,-}$ for $l$ being \emph{even} or $\omega_l = E_{m+l+1,-} - E_{m+l,+}$ for $l$ being \emph{odd}. Here, $\Omega$ relates to the driving amplitude and the above interaction can be regarded as a generalization of the so called ``all-resonant" control method \cite{Strauch2012}. In the interaction picture, the effective Hamiltonian reads
\begin{equation}
	H_{\rm I}^{(k)} = \hbar f_k(t) \sum_{j_1 \neq j_2}{}_{\mathrm{A},k}\langle j_1|\sigma_{x,k} |j_2\rangle_{\mathrm{A},k} 
	e^{\frac{i}{\hbar}(E_{j_1}-E_{j_2})t} |j_1\rangle_{\mathrm{A},k}{}_{\mathrm{A},k}\langle j_2|, \label{eq:hamiltonian_without_rwa}
\end{equation} 
where $|j\rangle_{\mathrm{A},k}$ represents the eigenstate corresponding to the eigenvalue $E_j$ for the system. Using the rotating-wave approximation, we find that the only non-zero elements of the effective Hamiltonian are 
\begin{equation}
	{}_{\mathrm{A},k}\langle m+l+1,\pm |H_{\rm I}| m+l,\mp \rangle_{\mathrm{A},k}
	= i\hbar \frac{\Omega}{2} \sqrt{l(2^k-l)}.
\end{equation}
This observation indicates that the driving field couples the states $|m+1,+\rangle_{\mathrm{A},k}$, $|m+2,-\rangle_{\mathrm{A},k}$, $\cdots$, $|m+2^k,-\rangle_{\mathrm{A},k}$ in a one-dimensional chain, and according to the results in perfect state transfer, the initial state $|m+1,+\rangle$ will be transferred to the final state $|m+2^k,-\rangle$ after time $\tau_{\rm map}=\pi/\Omega$ \cite{Christandl2004}. The initial state $|m,-\rangle$ will stay invariant since it is not coupled to any other states in the effective Hamiltonian. In total, the system evolves to the following state after time $\tau_{\rm map}$
\begin{equation}
	|\psi(\tau_{\rm map})\rangle_{\mathrm{A},k} 
	= \sum_{m} \left( c_0 |m,- \rangle_{\mathrm{A},k} + c_1 |m+2^k,-\rangle_{\mathrm{A},k} \right).
\end{equation}
Then, we adiabatically decouple the $k$th qubit and the resonator A and obtain 
\begin{equation}
	|\psi(\tau_{1})\rangle_{\mathrm{A},k} 
	= \sum_{m}\left( c_0 |m \rangle_{\rm A} + c_1 |m+2^k\rangle_{\rm A} \right)|0\rangle_{k},
\end{equation}
which means that the excitation of the $k$th qubit is transformed into additional $2^k$ photons in the resonator A.

We note that, the $k=0$ case requires a different driving field which is applied to the $0$th qubit
\begin{equation}
	V_{\rm map}^{(0)} = \sum_{m_0}\hbar 2\sqrt{2}\Omega 
	\left( -\cos(\omega_+ t) + \cos(\omega_1 t) \right) \sigma_{y,0}. \label{eq:drive_0}
\end{equation}
Here, $\omega_{+}=(E_{m_{0}+2,+} - E_{m_{0}+1,+})/\hbar$ and $\omega_1=(E_{m_{0}+2,+} - E_{m_{0}+1,-})/\hbar$. Following the same analysis above and under the assumption of the rotating wave approximation, the only non-zero elements of the effective Hamiltonian are 
\begin{eqnarray}
	{}_{\mathrm{A},k}\langle m+2,+ |H_{\rm I}| m+1,+ \rangle_{\mathrm{A},k}
	= -{}_{\mathrm{A},k}\langle m+1,+ |H_{\rm I}| m+2,+ \rangle_{\mathrm{A},k}
	= j\sqrt{2}, \\
	{}_{\mathrm{A},k}\langle m+1,- |H_{\rm I}| m+2,+ \rangle_{\mathrm{A},k}
	= -{}_{\mathrm{A},k}\langle m+2,+ |H_{\rm I}| m+1,- \rangle_{\mathrm{A},k}
	= j\sqrt{2}.
\end{eqnarray}
This means that the three states $|m+1,+\rangle$, $|m+2,+\rangle$, and $|m+1,-\rangle$ forms a perfect state transfer chain, which can be seen as a rotation of a fictitious spin-$1$ particle. Thus, the population of the initial state can be transferred to the target state after $\tau_{\rm map}$.

\end{widetext}

\bibliography{CavityQFT_submit}  

\begin{thebibliography}{44}%
\makeatletter
\providecommand \@ifxundefined [1]{%
 \@ifx{#1\undefined}
}%
\providecommand \@ifnum [1]{%
 \ifnum #1\expandafter \@firstoftwo
 \else \expandafter \@secondoftwo
 \fi
}%
\providecommand \@ifx [1]{%
 \ifx #1\expandafter \@firstoftwo
 \else \expandafter \@secondoftwo
 \fi
}%
\providecommand \natexlab [1]{#1}%
\providecommand \enquote  [1]{``#1''}%
\providecommand \bibnamefont  [1]{#1}%
\providecommand \bibfnamefont [1]{#1}%
\providecommand \citenamefont [1]{#1}%
\providecommand \href@noop [0]{\@secondoftwo}%
\providecommand \href [0]{\begingroup \@sanitize@url \@href}%
\providecommand \@href[1]{\@@startlink{#1}\@@href}%
\providecommand \@@href[1]{\endgroup#1\@@endlink}%
\providecommand \@sanitize@url [0]{\catcode `\\12\catcode `\$12\catcode
  `\&12\catcode `\#12\catcode `\^12\catcode `\_12\catcode `\%12\relax}%
\providecommand \@@startlink[1]{}%
\providecommand \@@endlink[0]{}%
\providecommand \url  [0]{\begingroup\@sanitize@url \@url }%
\providecommand \@url [1]{\endgroup\@href {#1}{\urlprefix }}%
\providecommand \urlprefix  [0]{URL }%
\providecommand \Eprint [0]{\href }%
\providecommand \doibase [0]{https://doi.org/}%
\providecommand \selectlanguage [0]{\@gobble}%
\providecommand \bibinfo  [0]{\@secondoftwo}%
\providecommand \bibfield  [0]{\@secondoftwo}%
\providecommand \translation [1]{[#1]}%
\providecommand \BibitemOpen [0]{}%
\providecommand \bibitemStop [0]{}%
\providecommand \bibitemNoStop [0]{.\EOS\space}%
\providecommand \EOS [0]{\spacefactor3000\relax}%
\providecommand \BibitemShut  [1]{\csname bibitem#1\endcsname}%
\let\auto@bib@innerbib\@empty
\bibitem [{\citenamefont {Neill}\ \emph {et~al.}(2018)\citenamefont {Neill},
  \citenamefont {Roushan}, \citenamefont {Kechedzhi}, \citenamefont {Boixo},
  \citenamefont {Isakov}, \citenamefont {Smelyanskiy}, \citenamefont {Megrant},
  \citenamefont {Chiaro}, \citenamefont {Dunsworth}, \citenamefont {Arya},
  \citenamefont {Barends}, \citenamefont {Burkett}, \citenamefont {Chen},
  \citenamefont {Chen}, \citenamefont {Fowler}, \citenamefont {Foxen},
  \citenamefont {Giustina}, \citenamefont {Graff}, \citenamefont {Jeffrey},
  \citenamefont {Huang}, \citenamefont {Kelly}, \citenamefont {Klimov},
  \citenamefont {Lucero}, \citenamefont {Mutus}, \citenamefont {Neeley},
  \citenamefont {Quintana}, \citenamefont {Sank}, \citenamefont {Vainsencher},
  \citenamefont {Wenner}, \citenamefont {White}, \citenamefont {Neven},\ and\
  \citenamefont {Martinis}}]{Neill2018}%
  \BibitemOpen
  \bibfield  {author} {\bibinfo {author} {\bibfnamefont {C.}~\bibnamefont
  {Neill}}, \bibinfo {author} {\bibfnamefont {P.}~\bibnamefont {Roushan}},
  \bibinfo {author} {\bibfnamefont {K.}~\bibnamefont {Kechedzhi}}, \bibinfo
  {author} {\bibfnamefont {S.}~\bibnamefont {Boixo}}, \bibinfo {author}
  {\bibfnamefont {S.~V.}\ \bibnamefont {Isakov}}, \bibinfo {author}
  {\bibfnamefont {V.}~\bibnamefont {Smelyanskiy}}, \bibinfo {author}
  {\bibfnamefont {A.}~\bibnamefont {Megrant}}, \bibinfo {author} {\bibfnamefont
  {B.}~\bibnamefont {Chiaro}}, \bibinfo {author} {\bibfnamefont
  {A.}~\bibnamefont {Dunsworth}}, \bibinfo {author} {\bibfnamefont
  {K.}~\bibnamefont {Arya}}, \bibinfo {author} {\bibfnamefont {R.}~\bibnamefont
  {Barends}}, \bibinfo {author} {\bibfnamefont {B.}~\bibnamefont {Burkett}},
  \bibinfo {author} {\bibfnamefont {Y.}~\bibnamefont {Chen}}, \bibinfo {author}
  {\bibfnamefont {Z.}~\bibnamefont {Chen}}, \bibinfo {author} {\bibfnamefont
  {A.}~\bibnamefont {Fowler}}, \bibinfo {author} {\bibfnamefont
  {B.}~\bibnamefont {Foxen}}, \bibinfo {author} {\bibfnamefont
  {M.}~\bibnamefont {Giustina}}, \bibinfo {author} {\bibfnamefont
  {R.}~\bibnamefont {Graff}}, \bibinfo {author} {\bibfnamefont
  {E.}~\bibnamefont {Jeffrey}}, \bibinfo {author} {\bibfnamefont
  {T.}~\bibnamefont {Huang}}, \bibinfo {author} {\bibfnamefont
  {J.}~\bibnamefont {Kelly}}, \bibinfo {author} {\bibfnamefont
  {P.}~\bibnamefont {Klimov}}, \bibinfo {author} {\bibfnamefont
  {E.}~\bibnamefont {Lucero}}, \bibinfo {author} {\bibfnamefont
  {J.}~\bibnamefont {Mutus}}, \bibinfo {author} {\bibfnamefont
  {M.}~\bibnamefont {Neeley}}, \bibinfo {author} {\bibfnamefont
  {C.}~\bibnamefont {Quintana}}, \bibinfo {author} {\bibfnamefont
  {D.}~\bibnamefont {Sank}}, \bibinfo {author} {\bibfnamefont {A.}~\bibnamefont
  {Vainsencher}}, \bibinfo {author} {\bibfnamefont {J.}~\bibnamefont {Wenner}},
  \bibinfo {author} {\bibfnamefont {T.~C.}\ \bibnamefont {White}}, \bibinfo
  {author} {\bibfnamefont {H.}~\bibnamefont {Neven}},\ and\ \bibinfo {author}
  {\bibfnamefont {J.~M.}\ \bibnamefont {Martinis}},\ }\bibfield  {title}
  {\bibinfo {title} {A blueprint for demonstrating quantum supremacy with
  superconducting qubits},\ }\href {https://doi.org/10.1126/science.aao4309}
  {\bibfield  {journal} {\bibinfo  {journal} {Science}\ }\textbf {\bibinfo
  {volume} {360}},\ \bibinfo {pages} {195} (\bibinfo {year}
  {2018})}\BibitemShut {NoStop}%
\bibitem [{\citenamefont {Arute}\ \emph {et~al.}(2019)\citenamefont {Arute},
  \citenamefont {Arya}, \citenamefont {Babbush}, \citenamefont {Bacon},
  \citenamefont {Bardin}, \citenamefont {Barends}, \citenamefont {Biswas},
  \citenamefont {Boixo}, \citenamefont {Brandao}, \citenamefont {Buell},
  \citenamefont {Burkett}, \citenamefont {Chen}, \citenamefont {Chen},
  \citenamefont {Chiaro}, \citenamefont {Collins}, \citenamefont {Courtney},
  \citenamefont {Dunsworth}, \citenamefont {Farhi}, \citenamefont {Foxen},
  \citenamefont {Fowler}, \citenamefont {Gidney}, \citenamefont {Giustina},
  \citenamefont {Graff}, \citenamefont {Guerin}, \citenamefont {Habegger},
  \citenamefont {Harrigan}, \citenamefont {Hartmann}, \citenamefont {Ho},
  \citenamefont {Hoffmann}, \citenamefont {Huang}, \citenamefont {Humble},
  \citenamefont {Isakov}, \citenamefont {Jeffrey}, \citenamefont {Jiang},
  \citenamefont {Kafri}, \citenamefont {Kechedzhi}, \citenamefont {Kelly},
  \citenamefont {Klimov}, \citenamefont {Knysh}, \citenamefont {Korotkov},
  \citenamefont {Kostritsa}, \citenamefont {Landhuis}, \citenamefont
  {Lindmark}, \citenamefont {Lucero}, \citenamefont {Lyakh}, \citenamefont
  {Mandrà}, \citenamefont {McClean}, \citenamefont {McEwen}, \citenamefont
  {Megrant}, \citenamefont {Mi}, \citenamefont {Michielsen}, \citenamefont
  {Mohseni}, \citenamefont {Mutus}, \citenamefont {Naaman}, \citenamefont
  {Neeley}, \citenamefont {Neill}, \citenamefont {Niu}, \citenamefont {Ostby},
  \citenamefont {Petukhov}, \citenamefont {Platt}, \citenamefont {Quintana},
  \citenamefont {Rieffel}, \citenamefont {Roushan}, \citenamefont {Rubin},
  \citenamefont {Sank}, \citenamefont {Satzinger}, \citenamefont {Smelyanskiy},
  \citenamefont {Sung}, \citenamefont {Trevithick}, \citenamefont
  {Vainsencher}, \citenamefont {Villalonga}, \citenamefont {White},
  \citenamefont {Yao}, \citenamefont {Yeh}, \citenamefont {Zalcman},
  \citenamefont {Neven},\ and\ \citenamefont {Martinis}}]{Arute2019}%
  \BibitemOpen
  \bibfield  {author} {\bibinfo {author} {\bibfnamefont {F.}~\bibnamefont
  {Arute}}, \bibinfo {author} {\bibfnamefont {K.}~\bibnamefont {Arya}},
  \bibinfo {author} {\bibfnamefont {R.}~\bibnamefont {Babbush}}, \bibinfo
  {author} {\bibfnamefont {D.}~\bibnamefont {Bacon}}, \bibinfo {author}
  {\bibfnamefont {J.~C.}\ \bibnamefont {Bardin}}, \bibinfo {author}
  {\bibfnamefont {R.}~\bibnamefont {Barends}}, \bibinfo {author} {\bibfnamefont
  {R.}~\bibnamefont {Biswas}}, \bibinfo {author} {\bibfnamefont
  {S.}~\bibnamefont {Boixo}}, \bibinfo {author} {\bibfnamefont {F.~G. S.~L.}\
  \bibnamefont {Brandao}}, \bibinfo {author} {\bibfnamefont {D.~A.}\
  \bibnamefont {Buell}}, \bibinfo {author} {\bibfnamefont {B.}~\bibnamefont
  {Burkett}}, \bibinfo {author} {\bibfnamefont {Y.}~\bibnamefont {Chen}},
  \bibinfo {author} {\bibfnamefont {Z.}~\bibnamefont {Chen}}, \bibinfo {author}
  {\bibfnamefont {B.}~\bibnamefont {Chiaro}}, \bibinfo {author} {\bibfnamefont
  {R.}~\bibnamefont {Collins}}, \bibinfo {author} {\bibfnamefont
  {W.}~\bibnamefont {Courtney}}, \bibinfo {author} {\bibfnamefont
  {A.}~\bibnamefont {Dunsworth}}, \bibinfo {author} {\bibfnamefont
  {E.}~\bibnamefont {Farhi}}, \bibinfo {author} {\bibfnamefont
  {B.}~\bibnamefont {Foxen}}, \bibinfo {author} {\bibfnamefont
  {A.}~\bibnamefont {Fowler}}, \bibinfo {author} {\bibfnamefont
  {C.}~\bibnamefont {Gidney}}, \bibinfo {author} {\bibfnamefont
  {M.}~\bibnamefont {Giustina}}, \bibinfo {author} {\bibfnamefont
  {R.}~\bibnamefont {Graff}}, \bibinfo {author} {\bibfnamefont
  {K.}~\bibnamefont {Guerin}}, \bibinfo {author} {\bibfnamefont
  {S.}~\bibnamefont {Habegger}}, \bibinfo {author} {\bibfnamefont {M.~P.}\
  \bibnamefont {Harrigan}}, \bibinfo {author} {\bibfnamefont {M.~J.}\
  \bibnamefont {Hartmann}}, \bibinfo {author} {\bibfnamefont {A.}~\bibnamefont
  {Ho}}, \bibinfo {author} {\bibfnamefont {M.}~\bibnamefont {Hoffmann}},
  \bibinfo {author} {\bibfnamefont {T.}~\bibnamefont {Huang}}, \bibinfo
  {author} {\bibfnamefont {T.~S.}\ \bibnamefont {Humble}}, \bibinfo {author}
  {\bibfnamefont {S.~V.}\ \bibnamefont {Isakov}}, \bibinfo {author}
  {\bibfnamefont {E.}~\bibnamefont {Jeffrey}}, \bibinfo {author} {\bibfnamefont
  {Z.}~\bibnamefont {Jiang}}, \bibinfo {author} {\bibfnamefont
  {D.}~\bibnamefont {Kafri}}, \bibinfo {author} {\bibfnamefont
  {K.}~\bibnamefont {Kechedzhi}}, \bibinfo {author} {\bibfnamefont
  {J.}~\bibnamefont {Kelly}}, \bibinfo {author} {\bibfnamefont {P.~V.}\
  \bibnamefont {Klimov}}, \bibinfo {author} {\bibfnamefont {S.}~\bibnamefont
  {Knysh}}, \bibinfo {author} {\bibfnamefont {A.}~\bibnamefont {Korotkov}},
  \bibinfo {author} {\bibfnamefont {F.}~\bibnamefont {Kostritsa}}, \bibinfo
  {author} {\bibfnamefont {D.}~\bibnamefont {Landhuis}}, \bibinfo {author}
  {\bibfnamefont {M.}~\bibnamefont {Lindmark}}, \bibinfo {author}
  {\bibfnamefont {E.}~\bibnamefont {Lucero}}, \bibinfo {author} {\bibfnamefont
  {D.}~\bibnamefont {Lyakh}}, \bibinfo {author} {\bibfnamefont
  {S.}~\bibnamefont {Mandrà}}, \bibinfo {author} {\bibfnamefont {J.~R.}\
  \bibnamefont {McClean}}, \bibinfo {author} {\bibfnamefont {M.}~\bibnamefont
  {McEwen}}, \bibinfo {author} {\bibfnamefont {A.}~\bibnamefont {Megrant}},
  \bibinfo {author} {\bibfnamefont {X.}~\bibnamefont {Mi}}, \bibinfo {author}
  {\bibfnamefont {K.}~\bibnamefont {Michielsen}}, \bibinfo {author}
  {\bibfnamefont {M.}~\bibnamefont {Mohseni}}, \bibinfo {author} {\bibfnamefont
  {J.}~\bibnamefont {Mutus}}, \bibinfo {author} {\bibfnamefont
  {O.}~\bibnamefont {Naaman}}, \bibinfo {author} {\bibfnamefont
  {M.}~\bibnamefont {Neeley}}, \bibinfo {author} {\bibfnamefont
  {C.}~\bibnamefont {Neill}}, \bibinfo {author} {\bibfnamefont {M.~Y.}\
  \bibnamefont {Niu}}, \bibinfo {author} {\bibfnamefont {E.}~\bibnamefont
  {Ostby}}, \bibinfo {author} {\bibfnamefont {A.}~\bibnamefont {Petukhov}},
  \bibinfo {author} {\bibfnamefont {J.~C.}\ \bibnamefont {Platt}}, \bibinfo
  {author} {\bibfnamefont {C.}~\bibnamefont {Quintana}}, \bibinfo {author}
  {\bibfnamefont {E.~G.}\ \bibnamefont {Rieffel}}, \bibinfo {author}
  {\bibfnamefont {P.}~\bibnamefont {Roushan}}, \bibinfo {author} {\bibfnamefont
  {N.~C.}\ \bibnamefont {Rubin}}, \bibinfo {author} {\bibfnamefont
  {D.}~\bibnamefont {Sank}}, \bibinfo {author} {\bibfnamefont {K.~J.}\
  \bibnamefont {Satzinger}}, \bibinfo {author} {\bibfnamefont {V.}~\bibnamefont
  {Smelyanskiy}}, \bibinfo {author} {\bibfnamefont {K.~J.}\ \bibnamefont
  {Sung}}, \bibinfo {author} {\bibfnamefont {M.~D.}\ \bibnamefont
  {Trevithick}}, \bibinfo {author} {\bibfnamefont {A.}~\bibnamefont
  {Vainsencher}}, \bibinfo {author} {\bibfnamefont {B.}~\bibnamefont
  {Villalonga}}, \bibinfo {author} {\bibfnamefont {T.}~\bibnamefont {White}},
  \bibinfo {author} {\bibfnamefont {Z.~J.}\ \bibnamefont {Yao}}, \bibinfo
  {author} {\bibfnamefont {P.}~\bibnamefont {Yeh}}, \bibinfo {author}
  {\bibfnamefont {A.}~\bibnamefont {Zalcman}}, \bibinfo {author} {\bibfnamefont
  {H.}~\bibnamefont {Neven}},\ and\ \bibinfo {author} {\bibfnamefont {J.~M.}\
  \bibnamefont {Martinis}},\ }\bibfield  {title} {\bibinfo {title} {Quantum
  supremacy using a programmable superconducting processor},\ }\href
  {https://doi.org/10.1038/s41586-019-1666-5} {\bibfield  {journal} {\bibinfo
  {journal} {Nature}\ }\textbf {\bibinfo {volume} {574}},\ \bibinfo {pages}
  {505} (\bibinfo {year} {2019})}\BibitemShut {NoStop}%
\bibitem [{\citenamefont {Nielsen}\ and\ \citenamefont
  {Chuang}(2010)}]{Nielsen2010}%
  \BibitemOpen
  \bibfield  {author} {\bibinfo {author} {\bibfnamefont {M.~A.}\ \bibnamefont
  {Nielsen}}\ and\ \bibinfo {author} {\bibfnamefont {I.~L.}\ \bibnamefont
  {Chuang}},\ }\href@noop {} {\emph {\bibinfo {title} {Quantum Computation and
  Quantum Information}}}\ (\bibinfo  {publisher} {Cambridge University Press},\
  \bibinfo {year} {2010})\BibitemShut {NoStop}%
\bibitem [{\citenamefont {{Shor}}(1994)}]{Shor1994}%
  \BibitemOpen
  \bibfield  {author} {\bibinfo {author} {\bibfnamefont {P.~W.}\ \bibnamefont
  {{Shor}}},\ }\bibfield  {title} {\bibinfo {title} {Algorithms for quantum
  computation: discrete logarithms and factoring},\ }in\ \href
  {https://doi.org/10.1109/SFCS.1994.365700} {\emph {\bibinfo {booktitle}
  {Proceedings 35th Annual Symposium on Foundations of Computer Science}}}\
  (\bibinfo {year} {1994})\ pp.\ \bibinfo {pages} {124--134}\BibitemShut
  {NoStop}%
\bibitem [{\citenamefont {Shor}(1999)}]{Shor1999}%
  \BibitemOpen
  \bibfield  {author} {\bibinfo {author} {\bibfnamefont {P.~W.}\ \bibnamefont
  {Shor}},\ }\bibfield  {title} {\bibinfo {title} {Polynomial-time algorithms
  for prime factorization and discrete logarithms on a quantum computer},\
  }\href {https://doi.org/10.1137/S0036144598347011} {\bibfield  {journal}
  {\bibinfo  {journal} {SIAM Rev.}\ }\textbf {\bibinfo {volume} {41}},\
  \bibinfo {pages} {303} (\bibinfo {year} {1999})}\BibitemShut {NoStop}%
\bibitem [{\citenamefont {Weinstein}\ \emph {et~al.}(2001)\citenamefont
  {Weinstein}, \citenamefont {Pravia}, \citenamefont {Fortunato}, \citenamefont
  {Lloyd},\ and\ \citenamefont {Cory}}]{Weinstein2001}%
  \BibitemOpen
  \bibfield  {author} {\bibinfo {author} {\bibfnamefont {Y.~S.}\ \bibnamefont
  {Weinstein}}, \bibinfo {author} {\bibfnamefont {M.~A.}\ \bibnamefont
  {Pravia}}, \bibinfo {author} {\bibfnamefont {E.~M.}\ \bibnamefont
  {Fortunato}}, \bibinfo {author} {\bibfnamefont {S.}~\bibnamefont {Lloyd}},\
  and\ \bibinfo {author} {\bibfnamefont {D.~G.}\ \bibnamefont {Cory}},\
  }\bibfield  {title} {\bibinfo {title} {Implementation of the quantum fourier
  transform},\ }\href {https://doi.org/10.1103/PhysRevLett.86.1889} {\bibfield
  {journal} {\bibinfo  {journal} {Phys. Rev. Lett.}\ }\textbf {\bibinfo
  {volume} {86}},\ \bibinfo {pages} {1889} (\bibinfo {year}
  {2001})}\BibitemShut {NoStop}%
\bibitem [{\citenamefont {Vandersypen}\ \emph {et~al.}(2001)\citenamefont
  {Vandersypen}, \citenamefont {Steffen}, \citenamefont {Breyta}, \citenamefont
  {Yannoni}, \citenamefont {Sherwood},\ and\ \citenamefont
  {Chuang}}]{Vandersypen2001}%
  \BibitemOpen
  \bibfield  {author} {\bibinfo {author} {\bibfnamefont {L.~M.~K.}\
  \bibnamefont {Vandersypen}}, \bibinfo {author} {\bibfnamefont
  {M.}~\bibnamefont {Steffen}}, \bibinfo {author} {\bibfnamefont
  {G.}~\bibnamefont {Breyta}}, \bibinfo {author} {\bibfnamefont {C.~S.}\
  \bibnamefont {Yannoni}}, \bibinfo {author} {\bibfnamefont {M.~H.}\
  \bibnamefont {Sherwood}},\ and\ \bibinfo {author} {\bibfnamefont {I.~L.}\
  \bibnamefont {Chuang}},\ }\bibfield  {title} {\bibinfo {title} {Experimental
  realization of shor's quantum factoring algorithm using nuclear magnetic
  resonance},\ }\href {https://doi.org/10.1038/414883a} {\bibfield  {journal}
  {\bibinfo  {journal} {Nature}\ }\textbf {\bibinfo {volume} {414}},\ \bibinfo
  {pages} {883} (\bibinfo {year} {2001})}\BibitemShut {NoStop}%
\bibitem [{\citenamefont {Politi}\ \emph {et~al.}(2009)\citenamefont {Politi},
  \citenamefont {Matthews},\ and\ \citenamefont
  {O{\textquoteright}Brien}}]{Politi2009}%
  \BibitemOpen
  \bibfield  {author} {\bibinfo {author} {\bibfnamefont {A.}~\bibnamefont
  {Politi}}, \bibinfo {author} {\bibfnamefont {J.~C.~F.}\ \bibnamefont
  {Matthews}},\ and\ \bibinfo {author} {\bibfnamefont {J.~L.}\ \bibnamefont
  {O{\textquoteright}Brien}},\ }\bibfield  {title} {\bibinfo {title}
  {Shor{\textquoteright}s quantum factoring algorithm on a photonic chip},\
  }\href {https://doi.org/10.1126/science.1173731} {\bibfield  {journal}
  {\bibinfo  {journal} {Science}\ }\textbf {\bibinfo {volume} {325}},\ \bibinfo
  {pages} {1221} (\bibinfo {year} {2009})}\BibitemShut {NoStop}%
\bibitem [{\citenamefont {Lucero}\ \emph {et~al.}(2012)\citenamefont {Lucero},
  \citenamefont {Barends}, \citenamefont {Chen}, \citenamefont {Kelly},
  \citenamefont {Mariantoni}, \citenamefont {Megrant}, \citenamefont
  {O’Malley}, \citenamefont {Sank}, \citenamefont {Vainsencher},
  \citenamefont {Wenner}, \citenamefont {White}, \citenamefont {Yin},
  \citenamefont {Cleland},\ and\ \citenamefont {Martinis}}]{Lucero2012}%
  \BibitemOpen
  \bibfield  {author} {\bibinfo {author} {\bibfnamefont {E.}~\bibnamefont
  {Lucero}}, \bibinfo {author} {\bibfnamefont {R.}~\bibnamefont {Barends}},
  \bibinfo {author} {\bibfnamefont {Y.}~\bibnamefont {Chen}}, \bibinfo {author}
  {\bibfnamefont {J.}~\bibnamefont {Kelly}}, \bibinfo {author} {\bibfnamefont
  {M.}~\bibnamefont {Mariantoni}}, \bibinfo {author} {\bibfnamefont
  {A.}~\bibnamefont {Megrant}}, \bibinfo {author} {\bibfnamefont
  {P.}~\bibnamefont {O’Malley}}, \bibinfo {author} {\bibfnamefont
  {D.}~\bibnamefont {Sank}}, \bibinfo {author} {\bibfnamefont {A.}~\bibnamefont
  {Vainsencher}}, \bibinfo {author} {\bibfnamefont {J.}~\bibnamefont {Wenner}},
  \bibinfo {author} {\bibfnamefont {T.}~\bibnamefont {White}}, \bibinfo
  {author} {\bibfnamefont {Y.}~\bibnamefont {Yin}}, \bibinfo {author}
  {\bibfnamefont {A.~N.}\ \bibnamefont {Cleland}},\ and\ \bibinfo {author}
  {\bibfnamefont {J.~M.}\ \bibnamefont {Martinis}},\ }\bibfield  {title}
  {\bibinfo {title} {Computing prime factors with a {J}osephson phase qubit
  quantum processor},\ }\href {https://doi.org/10.1038/nphys2385} {\bibfield
  {journal} {\bibinfo  {journal} {Nat. Phys.}\ }\textbf {\bibinfo {volume}
  {8}},\ \bibinfo {pages} {719} (\bibinfo {year} {2012})}\BibitemShut {NoStop}%
\bibitem [{\citenamefont {Cai}\ \emph {et~al.}(2013)\citenamefont {Cai},
  \citenamefont {Weedbrook}, \citenamefont {Su}, \citenamefont {Chen},
  \citenamefont {Gu}, \citenamefont {Zhu}, \citenamefont {Li}, \citenamefont
  {Liu}, \citenamefont {Lu},\ and\ \citenamefont {Pan}}]{Cai2013}%
  \BibitemOpen
  \bibfield  {author} {\bibinfo {author} {\bibfnamefont {X.~D.}\ \bibnamefont
  {Cai}}, \bibinfo {author} {\bibfnamefont {C.}~\bibnamefont {Weedbrook}},
  \bibinfo {author} {\bibfnamefont {Z.~E.}\ \bibnamefont {Su}}, \bibinfo
  {author} {\bibfnamefont {M.~C.}\ \bibnamefont {Chen}}, \bibinfo {author}
  {\bibfnamefont {M.}~\bibnamefont {Gu}}, \bibinfo {author} {\bibfnamefont
  {M.~J.}\ \bibnamefont {Zhu}}, \bibinfo {author} {\bibfnamefont
  {L.}~\bibnamefont {Li}}, \bibinfo {author} {\bibfnamefont {N.~L.}\
  \bibnamefont {Liu}}, \bibinfo {author} {\bibfnamefont {C.~Y.}\ \bibnamefont
  {Lu}},\ and\ \bibinfo {author} {\bibfnamefont {J.~W.}\ \bibnamefont {Pan}},\
  }\bibfield  {title} {\bibinfo {title} {Experimental quantum computing to
  solve systems of linear equations},\ }\href
  {https://doi.org/10.1103/PhysRevLett.110.230501} {\bibfield  {journal}
  {\bibinfo  {journal} {Phys. Rev. Lett.}\ }\textbf {\bibinfo {volume} {110}},\
  \bibinfo {pages} {230501} (\bibinfo {year} {2013})}\BibitemShut {NoStop}%
\bibitem [{\citenamefont {Pan}\ \emph {et~al.}(2014)\citenamefont {Pan},
  \citenamefont {Cao}, \citenamefont {Yao}, \citenamefont {Li}, \citenamefont
  {Ju}, \citenamefont {Chen}, \citenamefont {Peng}, \citenamefont {Kais},\ and\
  \citenamefont {Du}}]{Pan2014}%
  \BibitemOpen
  \bibfield  {author} {\bibinfo {author} {\bibfnamefont {J.}~\bibnamefont
  {Pan}}, \bibinfo {author} {\bibfnamefont {Y.}~\bibnamefont {Cao}}, \bibinfo
  {author} {\bibfnamefont {X.}~\bibnamefont {Yao}}, \bibinfo {author}
  {\bibfnamefont {Z.}~\bibnamefont {Li}}, \bibinfo {author} {\bibfnamefont
  {C.}~\bibnamefont {Ju}}, \bibinfo {author} {\bibfnamefont {H.}~\bibnamefont
  {Chen}}, \bibinfo {author} {\bibfnamefont {X.}~\bibnamefont {Peng}}, \bibinfo
  {author} {\bibfnamefont {S.}~\bibnamefont {Kais}},\ and\ \bibinfo {author}
  {\bibfnamefont {J.}~\bibnamefont {Du}},\ }\bibfield  {title} {\bibinfo
  {title} {Experimental realization of quantum algorithm for solving linear
  systems of equations},\ }\href {https://doi.org/10.1103/PhysRevA.89.022313}
  {\bibfield  {journal} {\bibinfo  {journal} {Phys. Rev. A}\ }\textbf {\bibinfo
  {volume} {89}},\ \bibinfo {pages} {022313} (\bibinfo {year}
  {2014})}\BibitemShut {NoStop}%
\bibitem [{\citenamefont {Zheng}\ \emph {et~al.}(2017)\citenamefont {Zheng},
  \citenamefont {Song}, \citenamefont {Chen}, \citenamefont {Xia},
  \citenamefont {Liu}, \citenamefont {Guo}, \citenamefont {Zhang},
  \citenamefont {Xu}, \citenamefont {Deng}, \citenamefont {Huang},
  \citenamefont {Wu}, \citenamefont {Yan}, \citenamefont {Zheng}, \citenamefont
  {Lu}, \citenamefont {Pan}, \citenamefont {Wang}, \citenamefont {Lu},\ and\
  \citenamefont {Zhu}}]{Zheng2017}%
  \BibitemOpen
  \bibfield  {author} {\bibinfo {author} {\bibfnamefont {Y.}~\bibnamefont
  {Zheng}}, \bibinfo {author} {\bibfnamefont {C.}~\bibnamefont {Song}},
  \bibinfo {author} {\bibfnamefont {M.-C.}\ \bibnamefont {Chen}}, \bibinfo
  {author} {\bibfnamefont {B.}~\bibnamefont {Xia}}, \bibinfo {author}
  {\bibfnamefont {W.}~\bibnamefont {Liu}}, \bibinfo {author} {\bibfnamefont
  {Q.}~\bibnamefont {Guo}}, \bibinfo {author} {\bibfnamefont {L.}~\bibnamefont
  {Zhang}}, \bibinfo {author} {\bibfnamefont {D.}~\bibnamefont {Xu}}, \bibinfo
  {author} {\bibfnamefont {H.}~\bibnamefont {Deng}}, \bibinfo {author}
  {\bibfnamefont {K.}~\bibnamefont {Huang}}, \bibinfo {author} {\bibfnamefont
  {Y.}~\bibnamefont {Wu}}, \bibinfo {author} {\bibfnamefont {Z.}~\bibnamefont
  {Yan}}, \bibinfo {author} {\bibfnamefont {D.}~\bibnamefont {Zheng}}, \bibinfo
  {author} {\bibfnamefont {L.}~\bibnamefont {Lu}}, \bibinfo {author}
  {\bibfnamefont {J.-W.}\ \bibnamefont {Pan}}, \bibinfo {author} {\bibfnamefont
  {H.}~\bibnamefont {Wang}}, \bibinfo {author} {\bibfnamefont {C.-Y.}\
  \bibnamefont {Lu}},\ and\ \bibinfo {author} {\bibfnamefont {X.}~\bibnamefont
  {Zhu}},\ }\bibfield  {title} {\bibinfo {title} {Solving systems of linear
  equations with a superconducting quantum processor},\ }\href
  {https://doi.org/10.1103/PhysRevLett.118.210504} {\bibfield  {journal}
  {\bibinfo  {journal} {Phys. Rev. Lett.}\ }\textbf {\bibinfo {volume} {118}},\
  \bibinfo {pages} {210504} (\bibinfo {year} {2017})}\BibitemShut {NoStop}%
\bibitem [{\citenamefont {Martin}\ \emph {et~al.}(2020)\citenamefont {Martin},
  \citenamefont {Lamata}, \citenamefont {Solano},\ and\ \citenamefont
  {Sanz}}]{Martin2020}%
  \BibitemOpen
  \bibfield  {author} {\bibinfo {author} {\bibfnamefont {A.}~\bibnamefont
  {Martin}}, \bibinfo {author} {\bibfnamefont {L.}~\bibnamefont {Lamata}},
  \bibinfo {author} {\bibfnamefont {E.}~\bibnamefont {Solano}},\ and\ \bibinfo
  {author} {\bibfnamefont {M.}~\bibnamefont {Sanz}},\ }\bibfield  {title}
  {\bibinfo {title} {Digital-analog quantum algorithm for the quantum fourier
  transform},\ }\href {https://doi.org/10.1103/PhysRevResearch.2.013012}
  {\bibfield  {journal} {\bibinfo  {journal} {Phys. Rev. Research}\ }\textbf
  {\bibinfo {volume} {2}},\ \bibinfo {pages} {013012} (\bibinfo {year}
  {2020})}\BibitemShut {NoStop}%
\bibitem [{\citenamefont {Griffiths}\ and\ \citenamefont
  {Niu}(1996)}]{Griffiths1996}%
  \BibitemOpen
  \bibfield  {author} {\bibinfo {author} {\bibfnamefont {R.~B.}\ \bibnamefont
  {Griffiths}}\ and\ \bibinfo {author} {\bibfnamefont {C.~S.}\ \bibnamefont
  {Niu}},\ }\bibfield  {title} {\bibinfo {title} {Semiclassical fourier
  transform for quantum computation},\ }\href
  {https://doi.org/10.1103/PhysRevLett.76.3228} {\bibfield  {journal} {\bibinfo
   {journal} {Phys. Rev. Lett.}\ }\textbf {\bibinfo {volume} {76}},\ \bibinfo
  {pages} {3228} (\bibinfo {year} {1996})}\BibitemShut {NoStop}%
\bibitem [{\citenamefont {Parker}\ and\ \citenamefont
  {Plenio}(2000)}]{Parker2000}%
  \BibitemOpen
  \bibfield  {author} {\bibinfo {author} {\bibfnamefont {S.}~\bibnamefont
  {Parker}}\ and\ \bibinfo {author} {\bibfnamefont {M.~B.}\ \bibnamefont
  {Plenio}},\ }\bibfield  {title} {\bibinfo {title} {Efficient factorization
  with a single pure qubit and $\mathrm{log}\mathit{N}$ mixed qubits},\ }\href
  {https://doi.org/10.1103/PhysRevLett.85.3049} {\bibfield  {journal} {\bibinfo
   {journal} {Phys. Rev. Lett.}\ }\textbf {\bibinfo {volume} {85}},\ \bibinfo
  {pages} {3049} (\bibinfo {year} {2000})}\BibitemShut {NoStop}%
\bibitem [{\citenamefont {Lu}\ \emph {et~al.}(2007)\citenamefont {Lu},
  \citenamefont {Browne}, \citenamefont {Yang},\ and\ \citenamefont
  {Pan}}]{Lu2007}%
  \BibitemOpen
  \bibfield  {author} {\bibinfo {author} {\bibfnamefont {C.~Y.}\ \bibnamefont
  {Lu}}, \bibinfo {author} {\bibfnamefont {D.~E.}\ \bibnamefont {Browne}},
  \bibinfo {author} {\bibfnamefont {T.}~\bibnamefont {Yang}},\ and\ \bibinfo
  {author} {\bibfnamefont {J.~W.}\ \bibnamefont {Pan}},\ }\bibfield  {title}
  {\bibinfo {title} {Demonstration of a compiled version of shor's quantum
  factoring algorithm using photonic qubits},\ }\href
  {https://doi.org/10.1103/PhysRevLett.99.250504} {\bibfield  {journal}
  {\bibinfo  {journal} {Phys. Rev. Lett.}\ }\textbf {\bibinfo {volume} {99}},\
  \bibinfo {pages} {250504} (\bibinfo {year} {2007})}\BibitemShut {NoStop}%
\bibitem [{\citenamefont {Lanyon}\ \emph {et~al.}(2007)\citenamefont {Lanyon},
  \citenamefont {Weinhold}, \citenamefont {Langford}, \citenamefont {Barbieri},
  \citenamefont {James}, \citenamefont {Gilchrist},\ and\ \citenamefont
  {White}}]{Lanyon2007}%
  \BibitemOpen
  \bibfield  {author} {\bibinfo {author} {\bibfnamefont {B.~P.}\ \bibnamefont
  {Lanyon}}, \bibinfo {author} {\bibfnamefont {T.~J.}\ \bibnamefont
  {Weinhold}}, \bibinfo {author} {\bibfnamefont {N.~K.}\ \bibnamefont
  {Langford}}, \bibinfo {author} {\bibfnamefont {M.}~\bibnamefont {Barbieri}},
  \bibinfo {author} {\bibfnamefont {D.~F.~V.}\ \bibnamefont {James}}, \bibinfo
  {author} {\bibfnamefont {A.}~\bibnamefont {Gilchrist}},\ and\ \bibinfo
  {author} {\bibfnamefont {A.~G.}\ \bibnamefont {White}},\ }\bibfield  {title}
  {\bibinfo {title} {Experimental demonstration of a compiled version of shor's
  algorithm with quantum entanglement},\ }\href
  {https://doi.org/10.1103/PhysRevLett.99.250505} {\bibfield  {journal}
  {\bibinfo  {journal} {Phys. Rev. Lett.}\ }\textbf {\bibinfo {volume} {99}},\
  \bibinfo {pages} {250505} (\bibinfo {year} {2007})}\BibitemShut {NoStop}%
\bibitem [{\citenamefont {Martin-Lopez}\ \emph {et~al.}(2012)\citenamefont
  {Martin-Lopez}, \citenamefont {Laing}, \citenamefont {Lawson}, \citenamefont
  {Alvarez}, \citenamefont {Zhou},\ and\ \citenamefont
  {O'brien}}]{Martin-Lopez2012}%
  \BibitemOpen
  \bibfield  {author} {\bibinfo {author} {\bibfnamefont {E.}~\bibnamefont
  {Martin-Lopez}}, \bibinfo {author} {\bibfnamefont {A.}~\bibnamefont {Laing}},
  \bibinfo {author} {\bibfnamefont {T.}~\bibnamefont {Lawson}}, \bibinfo
  {author} {\bibfnamefont {R.}~\bibnamefont {Alvarez}}, \bibinfo {author}
  {\bibfnamefont {X.~Q.}\ \bibnamefont {Zhou}},\ and\ \bibinfo {author}
  {\bibfnamefont {J.~L.}\ \bibnamefont {O'brien}},\ }\bibfield  {title}
  {\bibinfo {title} {Experimental realization of shor's quantum factoring
  algorithm using qubit recycling},\ }\href
  {https://doi.org/10.1038/nphoton.2012.259} {\bibfield  {journal} {\bibinfo
  {journal} {Nat. Photon.}\ }\textbf {\bibinfo {volume} {6}},\ \bibinfo {pages}
  {773} (\bibinfo {year} {2012})}\BibitemShut {NoStop}%
\bibitem [{\citenamefont {Zhou}\ \emph {et~al.}(2013)\citenamefont {Zhou},
  \citenamefont {Kalasuwan}, \citenamefont {Ralph},\ and\ \citenamefont
  {O'Brien}}]{Zhou2013}%
  \BibitemOpen
  \bibfield  {author} {\bibinfo {author} {\bibfnamefont {X.~Q.}\ \bibnamefont
  {Zhou}}, \bibinfo {author} {\bibfnamefont {P.}~\bibnamefont {Kalasuwan}},
  \bibinfo {author} {\bibfnamefont {T.~C.}\ \bibnamefont {Ralph}},\ and\
  \bibinfo {author} {\bibfnamefont {J.~L.}\ \bibnamefont {O'Brien}},\
  }\bibfield  {title} {\bibinfo {title} {Calculating unknown eigenvalues with a
  quantum algorithm},\ }\href {https://doi.org/10.1038/nphoton.2012.360}
  {\bibfield  {journal} {\bibinfo  {journal} {Nat. Photon.}\ }\textbf {\bibinfo
  {volume} {7}},\ \bibinfo {pages} {223} (\bibinfo {year} {2013})}\BibitemShut
  {NoStop}%
\bibitem [{\citenamefont {Smolin}\ \emph {et~al.}(2013)\citenamefont {Smolin},
  \citenamefont {Smith},\ and\ \citenamefont {Vargo}}]{Smolin2013}%
  \BibitemOpen
  \bibfield  {author} {\bibinfo {author} {\bibfnamefont {J.~A.}\ \bibnamefont
  {Smolin}}, \bibinfo {author} {\bibfnamefont {G.}~\bibnamefont {Smith}},\ and\
  \bibinfo {author} {\bibfnamefont {A.}~\bibnamefont {Vargo}},\ }\bibfield
  {title} {\bibinfo {title} {Oversimplifying quantum factoring},\ }\href
  {https://doi.org/10.1038/nature12290} {\bibfield  {journal} {\bibinfo
  {journal} {Nature}\ }\textbf {\bibinfo {volume} {499}},\ \bibinfo {pages}
  {163} (\bibinfo {year} {2013})}\BibitemShut {NoStop}%
\bibitem [{\citenamefont {Monz}\ \emph {et~al.}(2016)\citenamefont {Monz},
  \citenamefont {Nigg}, \citenamefont {Martinez}, \citenamefont {Brandl},
  \citenamefont {Schindler}, \citenamefont {Rines}, \citenamefont {Wang},
  \citenamefont {Chuang},\ and\ \citenamefont {Blatt}}]{Monz2016}%
  \BibitemOpen
  \bibfield  {author} {\bibinfo {author} {\bibfnamefont {T.}~\bibnamefont
  {Monz}}, \bibinfo {author} {\bibfnamefont {D.}~\bibnamefont {Nigg}}, \bibinfo
  {author} {\bibfnamefont {E.~A.}\ \bibnamefont {Martinez}}, \bibinfo {author}
  {\bibfnamefont {M.~F.}\ \bibnamefont {Brandl}}, \bibinfo {author}
  {\bibfnamefont {P.}~\bibnamefont {Schindler}}, \bibinfo {author}
  {\bibfnamefont {R.}~\bibnamefont {Rines}}, \bibinfo {author} {\bibfnamefont
  {S.~X.}\ \bibnamefont {Wang}}, \bibinfo {author} {\bibfnamefont {I.~L.}\
  \bibnamefont {Chuang}},\ and\ \bibinfo {author} {\bibfnamefont
  {R.}~\bibnamefont {Blatt}},\ }\bibfield  {title} {\bibinfo {title}
  {Realization of a scalable shor algorithm},\ }\href
  {https://doi.org/10.1126/science.aad9480} {\bibfield  {journal} {\bibinfo
  {journal} {Science}\ }\textbf {\bibinfo {volume} {351}},\ \bibinfo {pages}
  {1068} (\bibinfo {year} {2016})}\BibitemShut {NoStop}%
\bibitem [{\citenamefont {Harrow}\ \emph {et~al.}(2009)\citenamefont {Harrow},
  \citenamefont {Hassidim},\ and\ \citenamefont {Lloyd}}]{Harrow2009}%
  \BibitemOpen
  \bibfield  {author} {\bibinfo {author} {\bibfnamefont {A.~W.}\ \bibnamefont
  {Harrow}}, \bibinfo {author} {\bibfnamefont {A.}~\bibnamefont {Hassidim}},\
  and\ \bibinfo {author} {\bibfnamefont {S.}~\bibnamefont {Lloyd}},\ }\bibfield
   {title} {\bibinfo {title} {Quantum algorithm for linear systems of
  equations},\ }\href {https://doi.org/10.1103/PhysRevLett.103.150502}
  {\bibfield  {journal} {\bibinfo  {journal} {Phys. Rev. Lett.}\ }\textbf
  {\bibinfo {volume} {103}},\ \bibinfo {pages} {150502} (\bibinfo {year}
  {2009})}\BibitemShut {NoStop}%
\bibitem [{\citenamefont {Strauch}(2012)}]{Strauch2012}%
  \BibitemOpen
  \bibfield  {author} {\bibinfo {author} {\bibfnamefont {F.~W.}\ \bibnamefont
  {Strauch}},\ }\bibfield  {title} {\bibinfo {title} {All-resonant control of
  superconducting resonators},\ }\href
  {https://doi.org/10.1103/PhysRevLett.109.210501} {\bibfield  {journal}
  {\bibinfo  {journal} {Phys. Rev. Lett.}\ }\textbf {\bibinfo {volume} {109}},\
  \bibinfo {pages} {210501} (\bibinfo {year} {2012})}\BibitemShut {NoStop}%
\bibitem [{\citenamefont {Christandl}\ \emph {et~al.}(2004)\citenamefont
  {Christandl}, \citenamefont {Datta}, \citenamefont {Ekert},\ and\
  \citenamefont {Landahl}}]{Christandl2004}%
  \BibitemOpen
  \bibfield  {author} {\bibinfo {author} {\bibfnamefont {M.}~\bibnamefont
  {Christandl}}, \bibinfo {author} {\bibfnamefont {N.}~\bibnamefont {Datta}},
  \bibinfo {author} {\bibfnamefont {A.}~\bibnamefont {Ekert}},\ and\ \bibinfo
  {author} {\bibfnamefont {A.~J.}\ \bibnamefont {Landahl}},\ }\bibfield
  {title} {\bibinfo {title} {Perfect state transfer in quantum spin networks},\
  }\href {https://doi.org/10.1103/PhysRevLett.92.187902} {\bibfield  {journal}
  {\bibinfo  {journal} {Phys. Rev. Lett.}\ }\textbf {\bibinfo {volume} {92}},\
  \bibinfo {pages} {187902} (\bibinfo {year} {2004})}\BibitemShut {NoStop}%
\bibitem [{\citenamefont {Christandl}\ \emph {et~al.}(2005)\citenamefont
  {Christandl}, \citenamefont {Datta}, \citenamefont {Dorlas}, \citenamefont
  {Ekert}, \citenamefont {Kay},\ and\ \citenamefont
  {Landahl}}]{Christandl2005}%
  \BibitemOpen
  \bibfield  {author} {\bibinfo {author} {\bibfnamefont {M.}~\bibnamefont
  {Christandl}}, \bibinfo {author} {\bibfnamefont {N.}~\bibnamefont {Datta}},
  \bibinfo {author} {\bibfnamefont {T.~C.}\ \bibnamefont {Dorlas}}, \bibinfo
  {author} {\bibfnamefont {A.}~\bibnamefont {Ekert}}, \bibinfo {author}
  {\bibfnamefont {A.}~\bibnamefont {Kay}},\ and\ \bibinfo {author}
  {\bibfnamefont {A.~J.}\ \bibnamefont {Landahl}},\ }\bibfield  {title}
  {\bibinfo {title} {Perfect transfer of arbitrary states in quantum spin
  networks},\ }\href {https://doi.org/10.1103/PhysRevA.71.032312} {\bibfield
  {journal} {\bibinfo  {journal} {Phys. Rev. A}\ }\textbf {\bibinfo {volume}
  {71}},\ \bibinfo {pages} {032312} (\bibinfo {year} {2005})}\BibitemShut
  {NoStop}%
\bibitem [{\citenamefont {Axline}\ \emph {et~al.}(2016)\citenamefont {Axline},
  \citenamefont {Reagor}, \citenamefont {Heeres}, \citenamefont {Reinhold},
  \citenamefont {Wang}, \citenamefont {Shain}, \citenamefont {Pfaff},
  \citenamefont {Chu}, \citenamefont {Frunzio},\ and\ \citenamefont
  {Schoelkopf}}]{Axline2016}%
  \BibitemOpen
  \bibfield  {author} {\bibinfo {author} {\bibfnamefont {C.}~\bibnamefont
  {Axline}}, \bibinfo {author} {\bibfnamefont {M.}~\bibnamefont {Reagor}},
  \bibinfo {author} {\bibfnamefont {R.}~\bibnamefont {Heeres}}, \bibinfo
  {author} {\bibfnamefont {P.}~\bibnamefont {Reinhold}}, \bibinfo {author}
  {\bibfnamefont {C.}~\bibnamefont {Wang}}, \bibinfo {author} {\bibfnamefont
  {K.}~\bibnamefont {Shain}}, \bibinfo {author} {\bibfnamefont
  {W.}~\bibnamefont {Pfaff}}, \bibinfo {author} {\bibfnamefont
  {Y.}~\bibnamefont {Chu}}, \bibinfo {author} {\bibfnamefont {L.}~\bibnamefont
  {Frunzio}},\ and\ \bibinfo {author} {\bibfnamefont {R.~J.}\ \bibnamefont
  {Schoelkopf}},\ }\bibfield  {title} {\bibinfo {title} {An architecture for
  integrating planar and 3d cqed devices},\ }\href
  {https://doi.org/10.1063/1.4959241} {\bibfield  {journal} {\bibinfo
  {journal} {Appl. Phys. Lett.}\ }\textbf {\bibinfo {volume} {109}},\ \bibinfo
  {pages} {042601} (\bibinfo {year} {2016})}\BibitemShut {NoStop}%
\bibitem [{\citenamefont {Reagor}\ \emph {et~al.}(2013)\citenamefont {Reagor},
  \citenamefont {Paik}, \citenamefont {Catelani}, \citenamefont {Sun},
  \citenamefont {Axline}, \citenamefont {Holland}, \citenamefont {Pop},
  \citenamefont {Masluk}, \citenamefont {Brecht}, \citenamefont {Frunzio},
  \citenamefont {Devoret}, \citenamefont {Glazman},\ and\ \citenamefont
  {Schoelkopf}}]{Reagor2013}%
  \BibitemOpen
  \bibfield  {author} {\bibinfo {author} {\bibfnamefont {M.}~\bibnamefont
  {Reagor}}, \bibinfo {author} {\bibfnamefont {H.}~\bibnamefont {Paik}},
  \bibinfo {author} {\bibfnamefont {G.}~\bibnamefont {Catelani}}, \bibinfo
  {author} {\bibfnamefont {L.}~\bibnamefont {Sun}}, \bibinfo {author}
  {\bibfnamefont {C.}~\bibnamefont {Axline}}, \bibinfo {author} {\bibfnamefont
  {E.}~\bibnamefont {Holland}}, \bibinfo {author} {\bibfnamefont {I.~M.}\
  \bibnamefont {Pop}}, \bibinfo {author} {\bibfnamefont {N.~A.}\ \bibnamefont
  {Masluk}}, \bibinfo {author} {\bibfnamefont {T.}~\bibnamefont {Brecht}},
  \bibinfo {author} {\bibfnamefont {L.}~\bibnamefont {Frunzio}}, \bibinfo
  {author} {\bibfnamefont {M.~H.}\ \bibnamefont {Devoret}}, \bibinfo {author}
  {\bibfnamefont {L.}~\bibnamefont {Glazman}},\ and\ \bibinfo {author}
  {\bibfnamefont {R.~J.}\ \bibnamefont {Schoelkopf}},\ }\bibfield  {title}
  {\bibinfo {title} {Reaching 10 ms single photon lifetimes for superconducting
  aluminum cavities},\ }\href {https://doi.org/10.1063/1.4807015} {\bibfield
  {journal} {\bibinfo  {journal} {Appl. Phys. Lett.}\ }\textbf {\bibinfo
  {volume} {102}},\ \bibinfo {pages} {192604} (\bibinfo {year}
  {2013})}\BibitemShut {NoStop}%
\bibitem [{\citenamefont {Reagor}\ \emph {et~al.}(2016)\citenamefont {Reagor},
  \citenamefont {Pfaff}, \citenamefont {Axline}, \citenamefont {Heeres},
  \citenamefont {Ofek}, \citenamefont {Sliwa}, \citenamefont {Holland},
  \citenamefont {Wang}, \citenamefont {Blumoff}, \citenamefont {Chou},
  \citenamefont {Hatridge}, \citenamefont {Frunzio}, \citenamefont {Devoret},
  \citenamefont {Jiang},\ and\ \citenamefont {Schoelkopf}}]{Reagor2016}%
  \BibitemOpen
  \bibfield  {author} {\bibinfo {author} {\bibfnamefont {M.}~\bibnamefont
  {Reagor}}, \bibinfo {author} {\bibfnamefont {W.}~\bibnamefont {Pfaff}},
  \bibinfo {author} {\bibfnamefont {C.}~\bibnamefont {Axline}}, \bibinfo
  {author} {\bibfnamefont {R.~W.}\ \bibnamefont {Heeres}}, \bibinfo {author}
  {\bibfnamefont {N.}~\bibnamefont {Ofek}}, \bibinfo {author} {\bibfnamefont
  {K.}~\bibnamefont {Sliwa}}, \bibinfo {author} {\bibfnamefont
  {E.}~\bibnamefont {Holland}}, \bibinfo {author} {\bibfnamefont
  {C.}~\bibnamefont {Wang}}, \bibinfo {author} {\bibfnamefont {J.}~\bibnamefont
  {Blumoff}}, \bibinfo {author} {\bibfnamefont {K.}~\bibnamefont {Chou}},
  \bibinfo {author} {\bibfnamefont {M.~J.}\ \bibnamefont {Hatridge}}, \bibinfo
  {author} {\bibfnamefont {L.}~\bibnamefont {Frunzio}}, \bibinfo {author}
  {\bibfnamefont {M.~H.}\ \bibnamefont {Devoret}}, \bibinfo {author}
  {\bibfnamefont {L.}~\bibnamefont {Jiang}},\ and\ \bibinfo {author}
  {\bibfnamefont {R.~J.}\ \bibnamefont {Schoelkopf}},\ }\bibfield  {title}
  {\bibinfo {title} {Quantum memory with millisecond coherence in circuit
  qed},\ }\href {https://doi.org/10.1103/PhysRevB.94.014506} {\bibfield
  {journal} {\bibinfo  {journal} {Phys. Rev. B}\ }\textbf {\bibinfo {volume}
  {94}},\ \bibinfo {pages} {014506} (\bibinfo {year} {2016})}\BibitemShut
  {NoStop}%
\bibitem [{\citenamefont {Motzoi}\ and\ \citenamefont
  {Wilhelm}(2013)}]{Motzoi2013}%
  \BibitemOpen
  \bibfield  {author} {\bibinfo {author} {\bibfnamefont {F.}~\bibnamefont
  {Motzoi}}\ and\ \bibinfo {author} {\bibfnamefont {F.~K.}\ \bibnamefont
  {Wilhelm}},\ }\bibfield  {title} {\bibinfo {title} {{Improving frequency
  selection of driven pulses using derivative-based transition suppression}},\
  }\href {https://doi.org/10.1103/physreva.88.062318} {\bibfield  {journal}
  {\bibinfo  {journal} {Physical Review A}\ }\textbf {\bibinfo {volume} {88}},\
  \bibinfo {pages} {062318} (\bibinfo {year} {2013})}\BibitemShut {NoStop}%
\bibitem [{\citenamefont {Rebi\ifmmode~\acute{c}\else \'{c}\fi{}}\ \emph
  {et~al.}(2009)\citenamefont {Rebi\ifmmode~\acute{c}\else \'{c}\fi{}},
  \citenamefont {Twamley},\ and\ \citenamefont
  {Milburn}}]{Rebifmmodecuteclseci2009}%
  \BibitemOpen
  \bibfield  {author} {\bibinfo {author} {\bibfnamefont {S.}~\bibnamefont
  {Rebi\ifmmode~\acute{c}\else \'{c}\fi{}}}, \bibinfo {author} {\bibfnamefont
  {J.}~\bibnamefont {Twamley}},\ and\ \bibinfo {author} {\bibfnamefont {G.~J.}\
  \bibnamefont {Milburn}},\ }\bibfield  {title} {\bibinfo {title} {Giant kerr
  nonlinearities in circuit quantum electrodynamics},\ }\href
  {https://doi.org/10.1103/PhysRevLett.103.150503} {\bibfield  {journal}
  {\bibinfo  {journal} {Phys. Rev. Lett.}\ }\textbf {\bibinfo {volume} {103}},\
  \bibinfo {pages} {150503} (\bibinfo {year} {2009})}\BibitemShut {NoStop}%
\bibitem [{\citenamefont {Hu}\ \emph {et~al.}(2011)\citenamefont {Hu},
  \citenamefont {Ge}, \citenamefont {Chen}, \citenamefont {Yang},\ and\
  \citenamefont {Chen}}]{Hu2011}%
  \BibitemOpen
  \bibfield  {author} {\bibinfo {author} {\bibfnamefont {Y.}~\bibnamefont
  {Hu}}, \bibinfo {author} {\bibfnamefont {G.~Q.}\ \bibnamefont {Ge}}, \bibinfo
  {author} {\bibfnamefont {S.}~\bibnamefont {Chen}}, \bibinfo {author}
  {\bibfnamefont {X.~F.}\ \bibnamefont {Yang}},\ and\ \bibinfo {author}
  {\bibfnamefont {Y.~L.}\ \bibnamefont {Chen}},\ }\bibfield  {title} {\bibinfo
  {title} {Cross-kerr-effect induced by coupled josephson qubits in circuit
  quantum electrodynamics},\ }\href
  {https://doi.org/10.1103/PhysRevA.84.012329} {\bibfield  {journal} {\bibinfo
  {journal} {Phys. Rev. A}\ }\textbf {\bibinfo {volume} {84}},\ \bibinfo
  {pages} {012329} (\bibinfo {year} {2011})}\BibitemShut {NoStop}%
\bibitem [{\citenamefont {Liu}\ \emph {et~al.}(2004)\citenamefont {Liu},
  \citenamefont {Wei},\ and\ \citenamefont {Nori}}]{Liu2004}%
  \BibitemOpen
  \bibfield  {author} {\bibinfo {author} {\bibfnamefont {Y.~X.}\ \bibnamefont
  {Liu}}, \bibinfo {author} {\bibfnamefont {L.~F.}\ \bibnamefont {Wei}},\ and\
  \bibinfo {author} {\bibfnamefont {F.}~\bibnamefont {Nori}},\ }\bibfield
  {title} {\bibinfo {title} {Generation of nonclassical photon states using a
  superconducting qubit in a microcavity},\ }\href
  {http://stacks.iop.org/0295-5075/67/i=6/a=941} {\bibfield  {journal}
  {\bibinfo  {journal} {Europhys. Lett.}\ }\textbf {\bibinfo {volume} {67}},\
  \bibinfo {pages} {941} (\bibinfo {year} {2004})}\BibitemShut {NoStop}%
\bibitem [{\citenamefont {Hofheinz}\ \emph {et~al.}(2008)\citenamefont
  {Hofheinz}, \citenamefont {Weig}, \citenamefont {Ansmann}, \citenamefont
  {Bialczak}, \citenamefont {Lucero}, \citenamefont {Neeley}, \citenamefont
  {{AD}}, \citenamefont {Wang}, \citenamefont {Martinis},\ and\ \citenamefont
  {Cleland}}]{Hofheinz2008}%
  \BibitemOpen
  \bibfield  {author} {\bibinfo {author} {\bibfnamefont {M.}~\bibnamefont
  {Hofheinz}}, \bibinfo {author} {\bibfnamefont {E.}~\bibnamefont {Weig}},
  \bibinfo {author} {\bibfnamefont {M.}~\bibnamefont {Ansmann}}, \bibinfo
  {author} {\bibfnamefont {R.~C.}\ \bibnamefont {Bialczak}}, \bibinfo {author}
  {\bibfnamefont {E.}~\bibnamefont {Lucero}}, \bibinfo {author} {\bibfnamefont
  {M.}~\bibnamefont {Neeley}}, \bibinfo {author} {\bibfnamefont
  {O.}~\bibnamefont {{AD}}}, \bibinfo {author} {\bibfnamefont {H.}~\bibnamefont
  {Wang}}, \bibinfo {author} {\bibfnamefont {J.~M.}\ \bibnamefont {Martinis}},\
  and\ \bibinfo {author} {\bibfnamefont {A.}~\bibnamefont {Cleland}},\
  }\bibfield  {title} {\bibinfo {title} {Generation of fock states in a
  superconducting quantum circuit},\ }\href
  {https://doi.org/10.1038/nature07136} {\bibfield  {journal} {\bibinfo
  {journal} {Nature}\ }\textbf {\bibinfo {volume} {454}},\ \bibinfo {pages}
  {310} (\bibinfo {year} {2008})}\BibitemShut {NoStop}%
\bibitem [{\citenamefont {Hofheinz}\ \emph {et~al.}(2009)\citenamefont
  {Hofheinz}, \citenamefont {Wang}, \citenamefont {Ansmann}, \citenamefont
  {Bialczak}, \citenamefont {Lucero}, \citenamefont {Neeley}, \citenamefont
  {{AD}}, \citenamefont {Sank}, \citenamefont {Wenner}, \citenamefont
  {Martinis},\ and\ \citenamefont {Cleland}}]{Hofheinz2009}%
  \BibitemOpen
  \bibfield  {author} {\bibinfo {author} {\bibfnamefont {M.}~\bibnamefont
  {Hofheinz}}, \bibinfo {author} {\bibfnamefont {H.}~\bibnamefont {Wang}},
  \bibinfo {author} {\bibfnamefont {M.}~\bibnamefont {Ansmann}}, \bibinfo
  {author} {\bibfnamefont {R.~C.}\ \bibnamefont {Bialczak}}, \bibinfo {author}
  {\bibfnamefont {E.}~\bibnamefont {Lucero}}, \bibinfo {author} {\bibfnamefont
  {M.}~\bibnamefont {Neeley}}, \bibinfo {author} {\bibfnamefont
  {O.}~\bibnamefont {{AD}}}, \bibinfo {author} {\bibfnamefont {D.}~\bibnamefont
  {Sank}}, \bibinfo {author} {\bibfnamefont {J.}~\bibnamefont {Wenner}},
  \bibinfo {author} {\bibfnamefont {J.~M.}\ \bibnamefont {Martinis}},\ and\
  \bibinfo {author} {\bibfnamefont {A.}~\bibnamefont {Cleland}},\ }\bibfield
  {title} {\bibinfo {title} {Synthesizing arbitrary quantum states in a
  superconducting resonator},\ }\href {https://doi.org/10.1038/nature08005}
  {\bibfield  {journal} {\bibinfo  {journal} {Nature}\ }\textbf {\bibinfo
  {volume} {459}},\ \bibinfo {pages} {546} (\bibinfo {year}
  {2009})}\BibitemShut {NoStop}%
\bibitem [{\citenamefont {Strauch}\ \emph {et~al.}(2010)\citenamefont
  {Strauch}, \citenamefont {Jacobs},\ and\ \citenamefont
  {Simmonds}}]{Strauch2010}%
  \BibitemOpen
  \bibfield  {author} {\bibinfo {author} {\bibfnamefont {F.~W.}\ \bibnamefont
  {Strauch}}, \bibinfo {author} {\bibfnamefont {K.}~\bibnamefont {Jacobs}},\
  and\ \bibinfo {author} {\bibfnamefont {R.~W.}\ \bibnamefont {Simmonds}},\
  }\bibfield  {title} {\bibinfo {title} {Arbitrary control of entanglement
  between two superconducting resonators},\ }\href
  {https://doi.org/10.1103/PhysRevLett.105.050501} {\bibfield  {journal}
  {\bibinfo  {journal} {Phys. Rev. Lett.}\ }\textbf {\bibinfo {volume} {105}},\
  \bibinfo {pages} {050501} (\bibinfo {year} {2010})}\BibitemShut {NoStop}%
\bibitem [{\citenamefont {Strauch}\ \emph {et~al.}(2012)\citenamefont
  {Strauch}, \citenamefont {Onyango}, \citenamefont {Jacobs},\ and\
  \citenamefont {Simmonds}}]{Strauch2012a}%
  \BibitemOpen
  \bibfield  {author} {\bibinfo {author} {\bibfnamefont {F.~W.}\ \bibnamefont
  {Strauch}}, \bibinfo {author} {\bibfnamefont {D.}~\bibnamefont {Onyango}},
  \bibinfo {author} {\bibfnamefont {K.}~\bibnamefont {Jacobs}},\ and\ \bibinfo
  {author} {\bibfnamefont {R.~W.}\ \bibnamefont {Simmonds}},\ }\bibfield
  {title} {\bibinfo {title} {Entangled-state synthesis for superconducting
  resonators},\ }\href {https://doi.org/10.1103/PhysRevA.85.022335} {\bibfield
  {journal} {\bibinfo  {journal} {Phys. Rev. A}\ }\textbf {\bibinfo {volume}
  {85}},\ \bibinfo {pages} {022335} (\bibinfo {year} {2012})}\BibitemShut
  {NoStop}%
\bibitem [{\citenamefont {Sharma}\ and\ \citenamefont
  {Strauch}(2016)}]{Sharma2016}%
  \BibitemOpen
  \bibfield  {author} {\bibinfo {author} {\bibfnamefont {R.}~\bibnamefont
  {Sharma}}\ and\ \bibinfo {author} {\bibfnamefont {F.~W.}\ \bibnamefont
  {Strauch}},\ }\bibfield  {title} {\bibinfo {title} {Quantum state synthesis
  of superconducting resonators},\ }\href
  {https://doi.org/10.1103/PhysRevA.93.012342} {\bibfield  {journal} {\bibinfo
  {journal} {Phys. Rev. A}\ }\textbf {\bibinfo {volume} {93}},\ \bibinfo
  {pages} {012342} (\bibinfo {year} {2016})}\BibitemShut {NoStop}%
\bibitem [{\citenamefont {Wang}\ \emph {et~al.}(2017)\citenamefont {Wang},
  \citenamefont {Hu}, \citenamefont {Xu}, \citenamefont {Liu}, \citenamefont
  {Ma}, \citenamefont {Zheng}, \citenamefont {Vijay}, \citenamefont {Song},
  \citenamefont {Duan},\ and\ \citenamefont {Sun}}]{Wang2017}%
  \BibitemOpen
  \bibfield  {author} {\bibinfo {author} {\bibfnamefont {W.}~\bibnamefont
  {Wang}}, \bibinfo {author} {\bibfnamefont {L.}~\bibnamefont {Hu}}, \bibinfo
  {author} {\bibfnamefont {Y.}~\bibnamefont {Xu}}, \bibinfo {author}
  {\bibfnamefont {K.}~\bibnamefont {Liu}}, \bibinfo {author} {\bibfnamefont
  {Y.}~\bibnamefont {Ma}}, \bibinfo {author} {\bibfnamefont {S.~B.}\
  \bibnamefont {Zheng}}, \bibinfo {author} {\bibfnamefont {R.}~\bibnamefont
  {Vijay}}, \bibinfo {author} {\bibfnamefont {Y.~P.}\ \bibnamefont {Song}},
  \bibinfo {author} {\bibfnamefont {L.~M.}\ \bibnamefont {Duan}},\ and\
  \bibinfo {author} {\bibfnamefont {L.}~\bibnamefont {Sun}},\ }\bibfield
  {title} {\bibinfo {title} {Converting quasiclassical states into arbitrary
  fock state superpositions in a superconducting circuit},\ }\href
  {https://doi.org/10.1103/PhysRevLett.118.223604} {\bibfield  {journal}
  {\bibinfo  {journal} {Phys. Rev. Lett.}\ }\textbf {\bibinfo {volume} {118}},\
  \bibinfo {pages} {223604} (\bibinfo {year} {2017})}\BibitemShut {NoStop}%
\bibitem [{\citenamefont {Warren}\ \emph {et~al.}(1993)\citenamefont {Warren},
  \citenamefont {Rabitz},\ and\ \citenamefont {Dahleh}}]{Warren1993}%
  \BibitemOpen
  \bibfield  {author} {\bibinfo {author} {\bibfnamefont {W.~S.}\ \bibnamefont
  {Warren}}, \bibinfo {author} {\bibfnamefont {H.}~\bibnamefont {Rabitz}},\
  and\ \bibinfo {author} {\bibfnamefont {M.}~\bibnamefont {Dahleh}},\
  }\bibfield  {title} {\bibinfo {title} {Coherent control of quantum dynamics:
  The dream is alive},\ }\href {https://doi.org/10.1126/science.259.5101.1581}
  {\bibfield  {journal} {\bibinfo  {journal} {Science}\ }\textbf {\bibinfo
  {volume} {259}},\ \bibinfo {pages} {1581} (\bibinfo {year}
  {1993})}\BibitemShut {NoStop}%
\bibitem [{\citenamefont {Rabitz}\ \emph {et~al.}(2000)\citenamefont {Rabitz},
  \citenamefont {de~Vivie-Riedle}, \citenamefont {Motzkus},\ and\ \citenamefont
  {Kompa}}]{Rabitz2000}%
  \BibitemOpen
  \bibfield  {author} {\bibinfo {author} {\bibfnamefont {H.}~\bibnamefont
  {Rabitz}}, \bibinfo {author} {\bibfnamefont {R.}~\bibnamefont
  {de~Vivie-Riedle}}, \bibinfo {author} {\bibfnamefont {M.}~\bibnamefont
  {Motzkus}},\ and\ \bibinfo {author} {\bibfnamefont {K.}~\bibnamefont
  {Kompa}},\ }\bibfield  {title} {\bibinfo {title} {Whither the future of
  controlling quantum phenomena?},\ }\href
  {http://science.sciencemag.org/content/288/5467/824} {\bibfield  {journal}
  {\bibinfo  {journal} {Science}\ }\textbf {\bibinfo {volume} {288}},\ \bibinfo
  {pages} {824} (\bibinfo {year} {2000})}\BibitemShut {NoStop}%
\bibitem [{\citenamefont {Heeres}\ \emph {et~al.}(2017)\citenamefont {Heeres},
  \citenamefont {Reinhold}, \citenamefont {Ofek}, \citenamefont {Frunzio},
  \citenamefont {Jiang}, \citenamefont {Devoret},\ and\ \citenamefont
  {Schoelkopf}}]{Heeres2017}%
  \BibitemOpen
  \bibfield  {author} {\bibinfo {author} {\bibfnamefont {R.~W.}\ \bibnamefont
  {Heeres}}, \bibinfo {author} {\bibfnamefont {P.}~\bibnamefont {Reinhold}},
  \bibinfo {author} {\bibfnamefont {N.}~\bibnamefont {Ofek}}, \bibinfo {author}
  {\bibfnamefont {L.}~\bibnamefont {Frunzio}}, \bibinfo {author} {\bibfnamefont
  {L.}~\bibnamefont {Jiang}}, \bibinfo {author} {\bibfnamefont {M.~H.}\
  \bibnamefont {Devoret}},\ and\ \bibinfo {author} {\bibfnamefont {R.~J.}\
  \bibnamefont {Schoelkopf}},\ }\bibfield  {title} {\bibinfo {title}
  {Implementing a universal gate set on a logical qubit encoded in an
  oscillator},\ }\href {https://doi.org/10.1038/s41467-017-00045-1} {\bibfield
  {journal} {\bibinfo  {journal} {Nat. Commun.}\ }\textbf {\bibinfo {volume}
  {8}},\ \bibinfo {pages} {94} (\bibinfo {year} {2017})}\BibitemShut {NoStop}%
\bibitem [{\citenamefont {Hatridge}\ \emph {et~al.}(2013)\citenamefont
  {Hatridge}, \citenamefont {Shankar}, \citenamefont {Mirrahimi}, \citenamefont
  {Schackert}, \citenamefont {Geerlings}, \citenamefont {Brecht}, \citenamefont
  {Sliwa}, \citenamefont {Abdo}, \citenamefont {Frunzio}, \citenamefont
  {Girvin}, \citenamefont {Schoelkopf},\ and\ \citenamefont
  {Devoret}}]{Hatridge2013}%
  \BibitemOpen
  \bibfield  {author} {\bibinfo {author} {\bibfnamefont {M.}~\bibnamefont
  {Hatridge}}, \bibinfo {author} {\bibfnamefont {S.}~\bibnamefont {Shankar}},
  \bibinfo {author} {\bibfnamefont {M.}~\bibnamefont {Mirrahimi}}, \bibinfo
  {author} {\bibfnamefont {F.}~\bibnamefont {Schackert}}, \bibinfo {author}
  {\bibfnamefont {K.}~\bibnamefont {Geerlings}}, \bibinfo {author}
  {\bibfnamefont {T.}~\bibnamefont {Brecht}}, \bibinfo {author} {\bibfnamefont
  {K.~M.}\ \bibnamefont {Sliwa}}, \bibinfo {author} {\bibfnamefont
  {B.}~\bibnamefont {Abdo}}, \bibinfo {author} {\bibfnamefont {L.}~\bibnamefont
  {Frunzio}}, \bibinfo {author} {\bibfnamefont {S.~M.}\ \bibnamefont {Girvin}},
  \bibinfo {author} {\bibfnamefont {R.~J.}\ \bibnamefont {Schoelkopf}},\ and\
  \bibinfo {author} {\bibfnamefont {M.~H.}\ \bibnamefont {Devoret}},\
  }\bibfield  {title} {\bibinfo {title} {Quantum back-action of an individual
  variable-strength measurement},\ }\href
  {https://doi.org/10.1126/science.1226897} {\bibfield  {journal} {\bibinfo
  {journal} {Science}\ }\textbf {\bibinfo {volume} {339}},\ \bibinfo {pages}
  {178} (\bibinfo {year} {2013})}\BibitemShut {NoStop}%
\bibitem [{\citenamefont {Barenco}\ \emph {et~al.}(1996)\citenamefont
  {Barenco}, \citenamefont {Ekert}, \citenamefont {Suominen},\ and\
  \citenamefont {T\"orm\"a}}]{Barenco1996}%
  \BibitemOpen
  \bibfield  {author} {\bibinfo {author} {\bibfnamefont {A.}~\bibnamefont
  {Barenco}}, \bibinfo {author} {\bibfnamefont {A.}~\bibnamefont {Ekert}},
  \bibinfo {author} {\bibfnamefont {K.-A.}\ \bibnamefont {Suominen}},\ and\
  \bibinfo {author} {\bibfnamefont {P.}~\bibnamefont {T\"orm\"a}},\ }\bibfield
  {title} {\bibinfo {title} {Approximate quantum fourier transform and
  decoherence},\ }\href {https://doi.org/10.1103/PhysRevA.54.139} {\bibfield
  {journal} {\bibinfo  {journal} {Phys. Rev. A}\ }\textbf {\bibinfo {volume}
  {54}},\ \bibinfo {pages} {139} (\bibinfo {year} {1996})}\BibitemShut
  {NoStop}%
\bibitem [{\citenamefont {Gu\'ery-Odelin}\ \emph {et~al.}(2019)\citenamefont
  {Gu\'ery-Odelin}, \citenamefont {Ruschhaupt}, \citenamefont {Kiely},
  \citenamefont {Torrontegui}, \citenamefont {Mart\'{\i}nez-Garaot},\ and\
  \citenamefont {Muga}}]{GueryOdelin2019}%
  \BibitemOpen
  \bibfield  {author} {\bibinfo {author} {\bibfnamefont {D.}~\bibnamefont
  {Gu\'ery-Odelin}}, \bibinfo {author} {\bibfnamefont {A.}~\bibnamefont
  {Ruschhaupt}}, \bibinfo {author} {\bibfnamefont {A.}~\bibnamefont {Kiely}},
  \bibinfo {author} {\bibfnamefont {E.}~\bibnamefont {Torrontegui}}, \bibinfo
  {author} {\bibfnamefont {S.}~\bibnamefont {Mart\'{\i}nez-Garaot}},\ and\
  \bibinfo {author} {\bibfnamefont {J.~G.}\ \bibnamefont {Muga}},\ }\bibfield
  {title} {\bibinfo {title} {Shortcuts to adiabaticity: Concepts, methods, and
  applications},\ }\href {https://doi.org/10.1103/RevModPhys.91.045001}
  {\bibfield  {journal} {\bibinfo  {journal} {Rev. Mod. Phys.}\ }\textbf
  {\bibinfo {volume} {91}},\ \bibinfo {pages} {045001} (\bibinfo {year}
  {2019})}\BibitemShut {NoStop}%
\end{thebibliography}%
\end{document}